\documentclass[aps,prd,reprint,groupaddress,amsmath,amssymb]{revtex4-2}
\usepackage{color}
\usepackage{amsmath,amssymb}
\usepackage{graphicx}
\usepackage{ulem}
\usepackage[colorlinks, linkcolor=blue, citecolor=blue, urlcolor=blue, breaklinks=true]{hyperref}
\usepackage{subfig}
\usepackage{mathtools}

\DeclarePairedDelimiter\bra{\langle}{\rvert}
\DeclarePairedDelimiter\ket{\lvert}{\rangle}
\DeclarePairedDelimiterX\braket[2]{\langle}{\rangle}{#1\,\delimsize\vert\,\mathopen{}#2}

\begin{document}
	\title{Detectability of an orbital-angular-momentum-based gravitational wave detector}
	\author{Haorong Wu}
	\affiliation{School of Physics and Technology, Wuhan University, Wuhan 430072, China}
	\author{Xilong Fan}
	\email{xilong.fan@whu.edu.cn}
	\affiliation{School of Physics and Technology, Wuhan University, Wuhan 430072, China}
	
\begin{abstract}	
	The orbital angular momentum (OAM) of light provides a new approach to probing gravitational waves (GWs) through transitions between OAM states. Here, we assess the detectability of an OAM-based GW detector by developing a readout scheme and evaluating its noise budget over the frequency range from $1~\mathrm{mHz}$ to $100~\mathrm{Hz}$, including laser-intensity noise, laser-phase noise, mode-impurity noise, and shot noise. Whereas direct measurements of transition probabilities yield a response quadratic in the GW strain, interference with a reference OAM field produces an optical-power response linear in the strain. For a symmetric superposition OAM state, the GW-induced transition amplitudes satisfy $\alpha_+=-\alpha_-^*$, whereas the noise-induced mode-crosstalk coefficients obey $n_+=n_-^*$. Hence, the GW-induced signal forms a differential-mode contribution, while laser-intensity and mode-impurity noises enter as common-mode contributions, enabling their suppression by the balanced detection. Under a fixed beam-waist constraint, a shorter arm permits a larger paraxial parameter, reducing the equivalent angular-noise strain sensitivity. We adopt an arm length of $L=4\,\mathrm{m}$, $N=160$ round trips, a paraxial parameter of $\gamma=0.1$, a laser power incident on the beam splitter of $P_{\mathrm{BS}}=10\,\mathrm{kW}$, and a wavelength of $\lambda_{\mathrm{opt}}=1064\,\mathrm{nm}$. Using representative noise spectra, we find that, for GWs from an equal-mass binary black hole with source-frame component masses of $M_{\mathrm{src}}=10^{4}\,M_{\odot}$ at redshift $z=1$, balanced detection yields an SNR of $24.7$ for an imbalance parameter of $\epsilon=10^{-3}$ and $239$ for $\epsilon=10^{-5}$, with the sensitivity approaching the shot-noise limit in the latter case. These results highlight the importance of a compact optical configuration and strong common-mode rejection for realizing an OAM-based GW detector.
\end{abstract}
\maketitle

\section{Introduction}

Since the first direct detection of gravitational waves (GWs) by LIGO, the global detector network has recorded hundreds of GW signals \cite{abbott2016observation,PhysRevLett.119.161101,PhysRevX.13.041039}. GW detection efforts are concentrated in different frequency bands \cite{ni2018gravitational}. Ground-based laser interferometers probe the high frequency band, approximately $10~{\mathrm{Hz}}$--$100~{\mathrm{kHz}}$ \cite{abbott2016observation}, whereas space-based laser interferometers target the upper part of the low frequency band, approximately $10~\mu{\mathrm{Hz}}$--$0.1~{\mathrm{Hz}}$ \cite{amaro2012low}. At lower frequencies, pulsar timing arrays probe the very low frequency band, approximately $300~{\mathrm{pHz}}$--$100~{\mathrm{nHz}}$ \cite{hobbs2010international}, while cosmic microwave background experiments provide access to signatures associated with the extremely low frequency band, approximately $1~{\mathrm{aHz}}$--$10~{\mathrm{fHz}}$ \cite{wu2014guide}. In contrast, detection of GWs in the mid-frequency band, approximately $0.1~{\mathrm{Hz}}$--$10~{\mathrm{Hz}}$, remains challenging \cite{ni2018gravitational}. This motivates the exploration of additional optical degrees of freedom for GW detection. In particular, GW-induced transitions between the orbital angular momentum (OAM) of light provide an observable different from the optical phase and path-length measurements \cite{q38s-k2tq}. This opens the possibility of encoding GW information in transitions between OAM states rather than only in optical phase shifts.

Recently, we demonstrated that passing GWs can induce transitions in scalar vortex beams from an initial OAM state $\ket{l}$ to the sideband states $\ket{l\pm1}$ and $\ket{l\pm2}$ \cite{q38s-k2tq}. For parameters including a wavelength $\lambda_{\mathrm{opt}}=355~{\mathrm {nm}}$, radial wave number $k_\perp=1\times10^6~{\mathrm {m}}^{-1}$, incident laser power $P_{\mathrm {in}}=5500~{\mathrm {W}}$ enhanced by a power-recycling cavity, interaction length $L=1120~{\mathrm {km}}$, and GW amplitudes $A_+=A_\times=1\times10^{-21}$, these transitions can yield approximately $2880$ signal photons per second. However, our previous work established only the GW-induced transition amplitude $\alpha$, which scales linearly with the GW amplitude, $\alpha\propto A_\lambda$. The detectability of these transitions was not investigated. In particular, direct measurement of the GW-induced sideband power is challenging because the power scales as $A_\lambda^2\sim10^{-42}$ and can therefore be dominated by instrumental and optical noise. A quantitative noise budget and an appropriate readout scheme are thus required to assess the sensitivity of this method.

In this work, we enhance the readout of the GW-induced signal by employing a DC readout scheme in which the weak GW-induced state is superposed with a relatively strong reference state. The resulting interference term changes the signal scaling from $A_\lambda^2$ to $rA_\lambda$, where $r=0.01$ is the coefficient of the reference state. We then construct a noise budget over the frequency range $1~{\mathrm {mHz}}\leq f_g \leq100~{\mathrm {Hz}}$, including source-laser intensity noise, reference-laser intensity noise, reference-laser phase noise, angular noise, transverse displacement noise, mirror-surface-imperfection noise, and shot noise. Several of these noises do not modify the optical power but instead produce crosstalk into neighboring OAM states that overlap with the GW signal channels. To suppress such noise contributions, we employ balanced detection \cite{kumar2012versatile,Jia_2023,Bruynsteen:21,Jin:15,8444971}. With the initial photon state being a symmetric superposition state considered here, the technical noise contributions enter mainly through the common mode, whereas the GW-induced signal appears in the differential mode. Balanced detection therefore suppresses the common-mode noise while retaining the differential GW signal. We further demonstrate that a compact detector configuration can improve sensitivity by suppressing angular noise rather than enhancing the GW-induced signal. For a fixed beam waist, a shorter arm allows a larger paraxial parameter, reducing the equivalent strain noise associated with angular fluctuations while leaving the GW-induced transition amplitudes unchanged. Combined with balanced detection, this provides an effective approach to improving the detectability of the proposed detector. Last, we calculate the signal-to-noise ratio (SNR) for GWs emitted by compact binary systems with different masses and redshifts \cite{bilby_paper,bilby_doi,lalsuite,2022ApJ...935..167A,2018AJ....156..123A,2013A&A...558A..33A}. The results show that balanced detection significantly improves the detectability.

This paper is organized as follows. In Sec.~2, we review the interaction between photon OAM and GWs and introduce the configuration of an OAM-based GW detector. In Sec.~3, we formulate the noisy detector outputs and construct the noise budget, including source- and reference-laser intensity noise, angular noise, transverse displacement noise, mirror-surface-imperfection noise, and shot noise. In Sec.~4, we introduce balanced detection to suppress common-mode noise and derive the corresponding total sensitivity. In Sec.~5, we assess the detectability of the proposed detector by calculating the SNRs for GWs from compact binary systems with different source parameters. Finally, in Sec.~6, we summarize and discuss our results. Throughout this paper, we use natural units with $c=G=\hbar=1$.

\section{OAM-based GW detector}

\begin{figure}[tbhp]
	\centering
	\includegraphics[width=\linewidth]{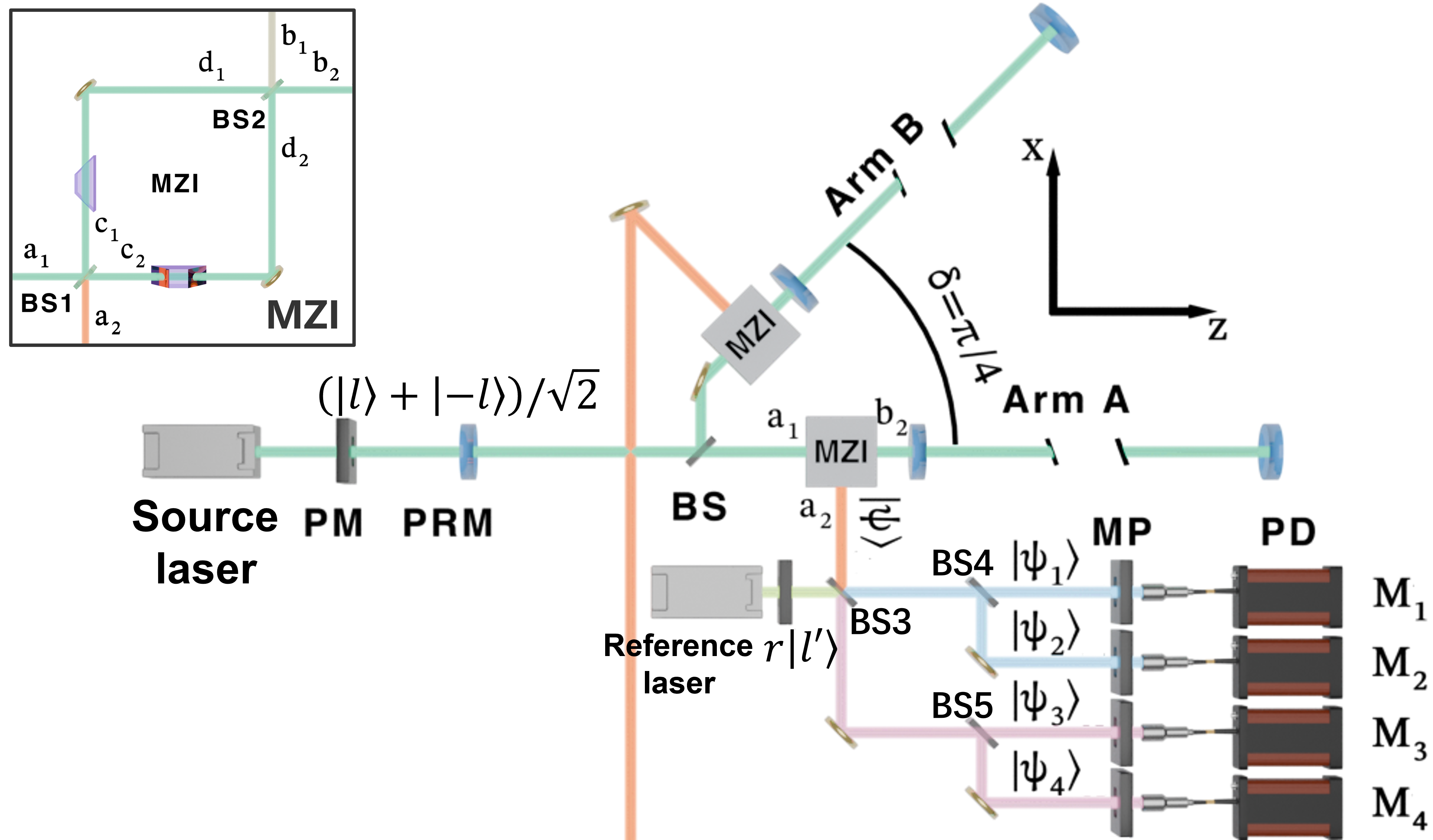}
	\caption{Schematic of the proposed GW detector. The detector comprises a phase modulator (PM), power-recycling mirror (PRM), beam splitter (BS), Mach--Zehnder interferometer (MZI), mode projectors (MPs), and photodetectors (PDs). Solid black squares indicate breaks in the optical paths and are used to compress the macroscopic arm lengths for illustration.}
	\label{fig.LightPath}
\end{figure}

In Ref.~\cite{wu2026displacementnoisefreegravitationalwavedetectorusing}, we proposed a V-shaped detector configuration, shown in Fig.~\ref{fig.LightPath}, that provides four real degrees of freedom for characterizing a GW signal. Because the underlying interaction and readout methods are identical for arms A and B, we restrict the analysis in this work to arm A. The results can be extended directly to the full two-arm configuration. The source laser is prepared by a phase modulator in the OAM superposition state $(\ket{l}+\ket{-l})/\sqrt{2}$, where $\ket{l}$ denotes a Bessel mode carrying OAM $l$ \cite{ARLT2000297,PhysRevLett.58.1499,PhysRevA.88.033818,PhysRevA.79.043802}. For propagation along the $z$ axis, the corresponding mode function is
\begin{equation}
	\mathrm{B}_{k_3 k_\perp l}(x^\alpha)=C_{k_3 k_\perp l}J_l(k_\perp\rho)e^{i(-Et+k_3z+l\phi)},
	\label{eq.BesselMode}
\end{equation}
where $C_{k_3 k_\perp l}=[k_\perp/(8 \pi^2 E)]^{1/2}$ is the normalization factor, $k_3$ is the longitudinal wave number, $k_\perp$ is the transverse wave number, $J_l(x)$ is the $l$th-order Bessel function of the first kind, and $E=(k^2_3+k^2_\perp)^{1/2}$. The light field is resonantly enhanced by a power-recycling cavity \cite{Aasi_2015}. In our previous work, the input field was prepared in a single OAM eigenstate $\ket{l}$. Here, we instead use the superposition $(\ket{l}+\ket{-l})/\sqrt{2}$ because, as shown later in the balanced-detection analysis, this state suppresses noise arising from mode impurity. The requirement $l\geq2$ follows from the same analysis and ensures separation between the two signal (or noise) channels. A beam splitter (BS) divides the optical power between arms A and B. We denote the power incident on the BS by $P_{\mathrm {BS}}$; for an ideal 50:50 BS, the power entering arm A is therefore $P_{\mathrm {BS}}/2$.

In arm A, a Mach--Zehnder interferometer (MZI) equipped with Dove prisms is used to separate different OAM states \cite{leach2002measuring,PhysRevLett.92.013601,PhysRevA.83.053829}. As illustrated in the inset of Fig.~\ref{fig.LightPath}, port $a_1$ is connected to the BS, port $a_2$ to BS3, and port $b_2$ to a Fabry--Perot cavity, while port $b_1$ is unused in the present configuration. Two Dove prisms are inserted into the internal paths $c_1$ and $c_2$. The Dove prism in path $c_1$ is rotated by $\phi=\pi/2$ about the optical axis, whereas that in path $c_2$ remains unrotated. The input state $(\ket{l}+\ket{-l})/\sqrt{2}$ from the BS enters through port $a_1$, while the vacuum state enters through port $a_2$. We choose the transformation matrix of BS1 as
\begin{equation}
	\frac{1}{\sqrt{2}}\begin{pmatrix}
		-1 & 1 \\
		1 & 1
	\end{pmatrix},
	\label{eq.TM1}
\end{equation}
where the factor $-1$ accounts for the $\pi$ phase shift acquired upon reflection at BS1 by the photons incident from port $a_1$. The state in the two internal paths is therefore
\begin{equation}
	\frac{1}{\sqrt{2}}
	\begin{pmatrix}
		-1 & 1 \\
		1 & 1
	\end{pmatrix}
	\begin{pmatrix}
		(\ket{l}+\ket{-l})/\sqrt{2} \\
		0
	\end{pmatrix}
	=
	\frac{1}{2}
	\begin{pmatrix}
		-\ket{l}-\ket{-l} \\
		\ket{l}+\ket{-l}
	\end{pmatrix}.
\end{equation}
The Dove prisms affect the OAM states in two ways \cite{leach2002measuring,PhysRevLett.92.013601,PhysRevA.83.053829}. First, a Dove prism reverses the OAM topological charge, $\ket{\pm l}\rightarrow\ket{\mp l}$. However, this reversal is compensated by the corresponding mirror reflections, so the net OAM topological charge is unchanged. Second, rotating a Dove prism by an angle $\phi$ about the optical axis introduces an OAM-dependent phase factor $e^{i2l\phi}$. For $\phi=\pi/2$, $\ket{l}\rightarrow e^{il\pi}\ket{l}
=(-1)^l\ket{l}$. The same factor is acquired by $\ket{-l}$ because $e^{-il\pi}=(-1)^l$. Hence, after the Dove prisms, the state becomes
\begin{equation}
	\frac{1}{2}
	\begin{pmatrix}
		(-1)^{l+1}(\ket{l}+\ket{-l}) \\
		\ket{l}+\ket{-l}
	\end{pmatrix}.
\end{equation}
The transformation matrix of BS2 is chosen as
\begin{equation}
	\frac{1}{\sqrt{2}}
	\begin{pmatrix}
		1 & 1 \\
		1 & -1
	\end{pmatrix}.
	\label{eq.TM2}
\end{equation}
The state immediately after BS2 is therefore
\begin{align}
	&
	\frac{1}{\sqrt{2}}
	\begin{pmatrix}
		1 & 1 \\
		1 & -1
	\end{pmatrix}
	\frac{1}{2}
	\begin{pmatrix}
		(-1)^{l+1}(\ket{l}+\ket{-l}) \\
		\ket{l}+\ket{-l}
	\end{pmatrix}
	\nonumber\\
	={}&
	\frac{1}{2\sqrt{2}}
	\begin{pmatrix}
		\left[(-1)^{l+1}+1\right](\ket{l}+\ket{-l}) \\
		\left[(-1)^{l+1}-1\right](\ket{l}+\ket{-l})
	\end{pmatrix}.
\end{align}
For even $l$, this reduces to $-(0,(\ket{l}+\ket{-l})/\sqrt 2)^{\text T}$, so the photons exit exclusively through port $b_2$. For odd $l$, they exit exclusively through port $b_1$. Without loss of generality, we take $l$ to be even. For odd $l$, interchanging the definitions of ports $b_1$ and $b_2$ yields the same subsequent analysis. The photons exiting through port $b_2$ then enter a Fabry--Perot cavity of length $L$ and complete $N$ round trips, during which they interact with the GW.

Suppose arm A is aligned with the $z$ axis, as shown in Fig.~\ref{fig.LightPath}, and the propagation direction of the GW is specified by the polar angle $\theta$ and azimuthal angle $\varphi$. Following Refs.~\cite{q38s-k2tq,wu2026displacementnoisefreegravitationalwavedetectorusing}, after interacting with the GW, the photon state becomes
\begin{align}
	-\frac{1}{\sqrt{2}}
	\Big[
	&\ket{l}+\ket{-l}
	+\alpha_{+}\left(\ket{l+1}+\ket{-l+1}\right)
	\nonumber\\
	&+\alpha_{-}\left(\ket{l-1}+\ket{-l-1}\right)
	\Big].
	\label{eq.AfterInteractionState}
\end{align}
Here, the transition amplitudes $\alpha_{\pm}$ are given by
\begin{equation}
	\alpha_{\pm}=\sum_{\lambda=+,\times}
	D_{\pm,\lambda}f_{\lambda}(t),
	\label{eq.alpha}
\end{equation}
where
\begin{align}
	f_{\lambda}(t)=&\frac{\sin(k_gNL)}{32k_g(x^2-1)}
	\Big[
	\cos(k_gNL-\Psi)
	-x\cot(k_gL)\nonumber\\
	&\times \sin(k_gNL-\Psi)
	-x\csc(k_gL)
	\sin(-Nk_gL\nonumber \\
	&+k_gLx+\Psi)
	\Big],
\end{align}
with $x=\cos\theta$ and $\Psi=\Psi_{\lambda}-k_gt$. The factors $D_{\pm,\lambda}$ are
\begin{align}
	D_{\pm,+}&=	\pm32\gamma k_3A_+\sin\theta\cos\theta e^{\pm i\varphi},\\
	D_{\pm,\times}&=32i\gamma k_3A_\times\sin\theta\,e^{\pm i\varphi},
\end{align}
where $\gamma=k_\perp/k_3$ is the paraxial parameter. The two transition amplitudes satisfy the symmetry relation
\begin{equation}
	\alpha_{+}=-\alpha_{-}^{*}.
\end{equation}

In the low-frequency regime, $k_gNL \ll 1$, the function $f_\lambda(t)$ can be simplified as
\begin{align}
	f_\lambda(t)
	=&
	-\frac{NL \cos(\Psi_\lambda-k_gt)}{32}
	\frac{\sin(k_gNL)}{k_gNL}
	\nonumber\\
	\approx &
	-\frac{NL \cos(\Psi_\lambda-k_gt)}{32}	
	\left[1+\left(\frac{f_g}{f_p}\right)^2\right]^{-1/2},
\end{align}
where $f_g=k_g/(2\pi)$ is the GW frequency and $f_p=1/(2\pi NL)$ is the cavity pole frequency. In the second line, we model the attenuation factor $\sin(k_gNL)/(k_gNL)$ by the single-pole response $[1+(f_g/f_p)^2]^{-1/2}$ \cite{9780198570745.001.0001}. The transition amplitude $\alpha_+$ can then be written as
\begin{equation}
	\alpha_+=	NL\gamma k_3
	\left[1+\left(\frac{f_g}{f_p}\right)^2\right]^{-1/2}
	h,
\end{equation}
where $h=h_{\mathrm {R}}+ih_{\mathrm {I}}$, with
\begin{align}
	h_{\mathrm {R}}=&\left (	\tilde A_\times\sin\varphi	-\tilde A_+\cos\theta\cos\varphi	\right)\sin\theta,\label{eq.hR}\\
	h_{\mathrm {I}}=&\left(-\tilde A_\times\cos\varphi-\tilde A_+\cos\theta\sin\varphi	\right)\sin\theta.\label{eq.hI}
\end{align}
Here, the effective GW amplitudes are defined as
\begin{equation}
	\tilde A_\lambda=A_\lambda\cos(\Psi_\lambda-k_gt).
\end{equation}
The two quantities $h_{\mathrm {R}}$ and $h_{\mathrm {I}}$ represent the real and imaginary quadratures of the effective GW response.

After interacting with the GW, the optical state $- [
\ket{l}+\ket{-l}
+\alpha_{+} (\ket{l+1}+\ket{-l+1} )
+\alpha_{-} (\ket{l-1}+\ket{-l-1} )
]/\sqrt{2}
$ enters the MZI through port $b_2$. Applying the BS2 transformation matrix in Eq.~\eqref{eq.TM2}, the state in the internal paths $d_1$ and $d_2$ becomes
\begin{widetext}
	\begin{equation}
		\frac{1}{2}
		\begin{pmatrix}
			-\left[
			\ket{l}+\ket{-l}
			+\alpha_{+}\left(\ket{l+1}+\ket{-l+1}\right)
			+\alpha_{-}\left(\ket{l-1}+\ket{-l-1}\right)
			\right]\\
			\ket{l}+\ket{-l}
			+\alpha_{+}\left(\ket{l+1}+\ket{-l+1}\right)
			+\alpha_{-}\left(\ket{l-1}+\ket{-l-1}\right)
		\end{pmatrix}.
	\end{equation}	
	As described above, the Dove prism in path $d_1$ is rotated by $\pi/2$ about the optical axis and therefore imparts the OAM-dependent phase factor $(-1)^m$ to an OAM state $\ket{m}$, whereas the unrotated Dove prism in path $d_2$ introduces no additional OAM-dependent phase. Since $l$ is chosen to be even, the states $\ket{\pm l}$ acquire a phase factor $+1$, while the GW-induced sideband states with OAM $l\pm1$ and $-l\pm1$ acquire a phase factor $-1$. Therefore, after passing through the Dove prisms, the state becomes
	\begin{equation}
		\frac{1}{2}
		\begin{pmatrix}
			-\ket{l}-\ket{-l}
			+\alpha_{+}\left(\ket{l+1}+\ket{-l+1}\right)
			+\alpha_{-}\left(\ket{l-1}+\ket{-l-1}\right)
			\\
			\ket{l}+\ket{-l}
			+\alpha_{+}\left(\ket{l+1}+\ket{-l+1}\right)
			+\alpha_{-}\left(\ket{l-1}+\ket{-l-1}\right)
		\end{pmatrix}.
	\end{equation}
\end{widetext}
Applying the BS1 transformation matrix in Eq.~\eqref{eq.TM1}, the final state at ports $a_1$ and $a_2$ is
\begin{equation}
	\frac 1{\sqrt 2}
	\begin{pmatrix}
		\ket{l}+\ket{-l}\\
		\alpha_{+}\left(\ket{l+1}+\ket{-l+1}\right)
		+\alpha_{-}\left(\ket{l-1}+\ket{-l-1}\right)
	\end{pmatrix}.
\end{equation}
Thus, the initial state $\left(\ket{l}+\ket{-l}\right)/\sqrt{2}$ exits exclusively through port $a_1$ and returns to the BS, whereas the GW-induced component,
\begin{equation}
	\ket{\psi}=\frac{\alpha_{+}\left(\ket{l+1}+\ket{-l+1}\right)
		+\alpha_{-}\left(\ket{l-1}+\ket{-l-1}\right)}{\sqrt{2}},
\end{equation}
exits exclusively through port $a_2$ and is directed to BS3. The MZI therefore spatially separates the GW-induced sideband states from the initial OAM state.

Since the transition amplitudes $\alpha_{\pm}$ scale linearly with the GW amplitude, $A_{\lambda}\sim10^{-21}$, direct measurements of the GW-induced states $\ket{l\pm1}$ and $\ket{-l\pm1}$ yield a signal proportional to $A_{\lambda}^{2} \sim 10^{-42}$. For example, with an input power of $5.5~{\mathrm {kW}}$ and a total propagation distance of $1120~{\mathrm {km}}$, the GW-induced transitions produce only approximately $2880$ signal photons per second \cite{q38s-k2tq}. To enhance the readout, the GW-induced state $\ket{\psi}$ is superposed with a relatively strong reference light field in an OAM state $\ket{l'}$. The choice of $l'$ must satisfy two conditions. First, if $l'$ coincides with one of the GW-induced sideband states, namely $l+1$, $l-1$, $-l+1$, or $-l-1$, the reference state can dominate the corresponding GW-induced component. Second, because imperfections in the MZI may cause a fraction of the initial state $(\ket{l}+\ket{-l})/\sqrt{2}$ to leak into port $a_2$, the reference state should also be different from $\ket{l}$ and $\ket{-l}$ to avoid overlap with this leakage. Therefore, we require
$l' \notin \{l,-l,l+1,l-1,-l+1,-l-1\}$.

The reference state is provided by the reference laser with power $r^2P_{\mathrm {BS}}/2$, where $r$ is a small number. To ensure stable interference with the GW-induced state, the reference laser is assumed to be phase locked with the source laser. The light from the MZI and the reference laser are combined at BS3. Assuming an ideal 50:50 splitting ratio, the transformation matrix of BS3 is
\begin{equation}
	\frac{1}{\sqrt{2}}
	\begin{pmatrix}
		1 & 1 \\
		1 & -1
	\end{pmatrix}.
\end{equation}
The two output states from BS3 are therefore
\begin{equation}
	\frac{1}{\sqrt{2}}
	\begin{pmatrix}
		1 & 1 \\
		1 & -1
	\end{pmatrix}
	\begin{pmatrix}
		\ket{\psi} \\
		r\ket{l'}
	\end{pmatrix}=
	\frac{1}{\sqrt{2}}
	\begin{pmatrix}
		\ket{\psi}+r\ket{l'} \\
		\ket{\psi}-r\ket{l'}
	\end{pmatrix}.
\end{equation}
As illustrated in Fig.~\ref{fig.LightPath}, the two outputs from BS3 are subsequently divided equally by BS4 and BS5, respectively. For BS4, we obtain
\begin{equation}
	\frac{1}{\sqrt{2}}
	\begin{pmatrix}
		1 & 1 \\
		1 & -1
	\end{pmatrix}
	\frac{1}{\sqrt{2}}
	\begin{pmatrix}
		\ket{\psi}+r\ket{l'} \\
		0
	\end{pmatrix}=	\begin{pmatrix}
		\ket{\psi_1} \\
		\ket{\psi_2}
	\end{pmatrix},
\end{equation}
whereas for BS5,
\begin{equation}
	\frac{1}{\sqrt{2}}
	\begin{pmatrix}
		1 & 1 \\
		1 & -1
	\end{pmatrix}
	\frac{1}{\sqrt{2}}
	\begin{pmatrix}
		\ket{\psi}-r\ket{l'} \\
		0
	\end{pmatrix}=
	\begin{pmatrix}
		\ket{\psi_3} \\
		\ket{\psi_4}
	\end{pmatrix}.
\end{equation}
Hence,
\begin{align}
	\ket{\psi_1}=\ket{\psi_2}=&\frac{\ket{\psi}+r\ket{l'}}{2},\\
	\ket{\psi_3}=\ket{\psi_4}=&\frac{\ket{\psi}-r\ket{l'}}{2}.
\end{align}

To measure the complex transition amplitudes $\alpha_{\pm}$, the four output states $\ket{\psi_m}$, with $m=1,\ldots,4$, are projected onto the corresponding states
\begin{align}
	\ket{\Phi_1}=&\frac{\sqrt{2}\ket{l'}+\ket{l+1}+\ket{-l+1}}{2},\\
	\ket{\Phi_2}=&\frac{\sqrt{2}\ket{l'}+i\ket{l+1}+i\ket{-l+1}}{2},\\
	\ket{\Phi_3}=&\frac{-\sqrt{2}\ket{l'}+\ket{l-1}+\ket{-l-1}}{2},	\\
	\ket{\Phi_4}=&\frac{-\sqrt{2}\ket{l'}-i\ket{l-1}-i\ket{-l-1}}{2}.
\end{align}
The power of corresponding projection measurements are
\begin{align}
	M_1=&\frac{P_{\mathrm {BS}}}{2}\left|\braket{\Phi_1}{\psi_1}\right|^2,\label{eq.MP1}\\
	M_2=&\frac{P_{\mathrm {BS}}}{2}\left|\braket{\Phi_2}{\psi_2}\right|^2,\\
	M_3=&\frac{P_{\mathrm {BS}}}{2}\left|\braket{\Phi_3}{\psi_3}\right|^2,\\
	M_4=&\frac{P_{\mathrm {BS}}}{2}\left|\braket{\Phi_4}{\psi_4}\right|^2.\label{eq.MP4}
\end{align}
Because $\alpha_{\pm}$ are first order in the GW amplitude, we retain only terms up to first order in $\alpha_{\pm}$. For example, Eq.~\eqref{eq.MP1} gives
\begin{align}
	M_1=&\frac{P_{\mathrm {BS}}}{2}\left|\braket{\Phi_1}{\psi_1}\right|^2\nonumber\\
	=&\frac{P_{\mathrm {BS}}}{2}\bigg|\frac{\sqrt{2}\bra{l'}+\bra{l+1}+\bra{-l+1}}{2}
	\frac{1}{2\sqrt{2}}	\Big[	\sqrt{2}r\ket{l'}
	+\alpha_+\nonumber \\
	&\times \left(\ket{l+1}+\ket{-l+1}\right)+\alpha_-\left(\ket{l-1}+\ket{-l-1}\right)
	\Big]	\bigg|^2\nonumber\\
	=&\frac{P_{\mathrm {BS}}}{2}\left|\frac{r+\alpha_+}{2\sqrt{2}}\right|^2\nonumber\\
	=&\frac{P_{\mathrm {BS}}}{2}\left[\frac{r^2}{8}+\frac{r\operatorname{Re}(\alpha_+)}{4}	+\frac{|\alpha_+|^2}{8}	\right]	\nonumber\\
	\approx&\frac{P_{\mathrm {BS}}}{2}\left[\frac{r^2}{8}+\frac{r\operatorname{Re}(\alpha_+)}{4}
	\right].\label{eq.M1}
\end{align}
Similarly, the remaining projection measurements are
\begin{align}
	M_2\approx&\frac{P_{\mathrm {BS}}}{2}\left[\frac{r^2}{8}+\frac{r\operatorname{Im}(\alpha_+)}{4}
	\right],\\
	M_3\approx&\frac{P_{\mathrm {BS}}}{2}\left[\frac{r^2}{8}+\frac{r\operatorname{Re}(\alpha_-)}{4}
	\right],\\
	M_4\approx&\frac{P_{\mathrm {BS}}}{2}\left[\frac{r^2}{8}-\frac{r\operatorname{Im}(\alpha_-)}{4}
	\right].
\end{align}

In this paper, we choose the laser wavelength to be $\lambda_{\mathrm{opt}}=1064~{\mathrm {nm}}$ and the laser power incident on the BS after enhancement by the power-recycling cavity to be $P_{\mathrm {BS}}=10~{\mathrm {kW}}$. The coefficient of the reference state $\ket{l'}$ is set to $r=0.01$. We consider a short-arm configuration, with $L=4~{\mathrm {m}}$, and $N$ is set to $160$. It is worth noting that the interaction theory in Eq.~\eqref{eq.BesselMode} is formulated using ideal Bessel modes, which require infinite energy and are therefore not physically realizable. In practice, we can only generate Bessel--Gaussian modes. Their mode function can be written as
\begin{align}
	&{\mathrm {BG}}_{k_3 k_\perp l}(x^\alpha)\nonumber \\
	=&
	\frac{C_{k_3 k_\perp l}w_0}{w(z)}
	J_l\left(\frac{k_\perp\rho}{1+iz/z_R}\right)
	\exp\bigg \{\left[\frac{-1}{w^2(z)}+\frac{ik}{2R(z)}\right]\nonumber \\
	&\times 
		\left (
		\rho^2+\frac{k_\perp^2z^2}{k^2}\right )\bigg \}
	\exp\bigg \{	i\bigg [	\left(	k-\frac{k_\perp^2}{2k}
		\right)z-\psi_G(z)\nonumber \\
		&-Et+l\phi \bigg ]\bigg \},
	\label{eq.BGMode}
\end{align}
where $k=(k^2_3+k^2_\perp)^{1/2}$, $w_0$ is the waist radius, $w(z)=w_0(1+z^2/z_R^2)^{1/2}$, $z_R=k_3w_0^2/2$ is the Rayleigh range, $R(z)=z (1+z_R^2/z^2 )$ is the radius of curvature, and $\psi_G(z)=\arctan (z/z_R ) $ is the Gouy phase. A Bessel--Gaussian beam has a finite diffraction-free range, $z_{\mathrm {max}}=w_0/\gamma$ \cite{PhysRevLett.58.1499,Liu:26}. Within this range, its transverse intensity profile remains approximately constant, whereas beyond this range the profile progressively deviates from that of an ideal Bessel mode. By choosing $L=z_{\mathrm {max}}$ and a waist radius $w_0=0.4~{\mathrm {m}}$, we obtain $\gamma=0.1$ for $L=4~{\mathrm {m}}$.

\section{Noise budget analysis}
We analyze the noise budget of the detector over the frequency range $1~{\mathrm {mHz}}$ to $100~{\mathrm {Hz}}$, including laser-intensity noise, laser-phase noise, mode-impurity noise, and shot noise. This interval covers a broad range of compact-binary signals. The lower-frequency portion is also particularly relevant for assessing technical noises such as angular and transverse-displacement fluctuations, which become increasingly important toward low frequencies. We choose $100~\mathrm{Hz}$ as the upper limit because it is sufficient to capture the transition from the low-frequency technical-noise-dominated regime to the high-frequency shot-noise-limited regime considered here. Moreover, this frequency remains well below the pole frequency $f_p=1/(2\pi NL)=7.46\times10^4~\mathrm{Hz}$ for $L=4~\mathrm{m}$, so the GW-induced response is essentially unaffected by cavity attenuation throughout the frequency band. This frequency interval is not a fundamental limit of the proposed detector and can, in principle, be extended further, particularly toward lower frequencies, provided that the corresponding low-frequency technical noises can be sufficiently characterized. We begin by deriving the noisy detector outputs.

\subsection{Noisy outputs}

We take the output $M_1$ as an example. The dominant technical noises are modeled as small perturbations to the parameters and mode amplitudes entering $M_1$, while shot noise is included as an additive contribution to the measured output. From the second line of Eq.~\eqref{eq.M1}, the noisy output can be written as
\begin{align}
	\widetilde{M}_1
	=&\frac{P_{\mathrm {BS}}(1+\delta_{\mathrm {SL}})}{2}
	\Bigg|
	\bra{\Phi_1}
	\frac{1}{2\sqrt{2}}
	\Big[
	\sqrt{2}r(1+\delta_r)e^{i\delta_{\phi_r}}\ket{l'}
	+(\alpha_+\nonumber \\
	&+n_{l,+})\ket{l+1}
	+(\alpha_-+n_{l,-})\ket{l-1}
	+(\alpha_++n_{-l,+})\nonumber \\
	&\times \ket{-l+1}+(\alpha_-+n_{-l,-})\ket{-l-1}\Big ]\Bigg |^2	+\delta_{1,{\mathrm {shot}}}
	\nonumber\\
	\approx&
	\frac{r^2P_{\mathrm {BS}}}{16}
	+\frac{r \operatorname{Re} (\alpha_+)P_{\mathrm {BS}}}{8}
	+\frac{r^2\delta_{\mathrm {SL}}P_{\mathrm {BS}}}{16}
	+\frac{r^2\delta_r P_{\mathrm {BS}}}{8}
	\nonumber\\
	&+
	\frac{r \operatorname{Re} (n_+)P_{\mathrm {BS}}}{16}
	+\delta_{1,{\mathrm {shot}}}
	\nonumber\\
	=&
	P_{1,{\mathrm {DC}}}
	+S_{1,{\mathrm {GW}}}
	+N_{1,{\mathrm {SL}}}
	+N_{1,r}
	+N_{1,{\mathrm {mode}}}
	+N_{1,{\mathrm {shot}}},
	\label{eq.noisyM1}
\end{align}
where $n_{\pm}=n_{l,\pm}+n_{-l,\pm}$. Here, $P_{1,\mathrm {DC}}=r^2P_{\mathrm {BS}}/16$ and $S_{1,\mathrm {GW}}=r\operatorname{Re}(\alpha_+)P_{\mathrm {BS}}/8$ are the DC output and the GW-induced signal, respectively. The corresponding noise contributions are $N_{1,{\mathrm {SL}}}=r^2\delta_{\mathrm {SL}}P_{\mathrm{BS}}/16$, $N_{1,r}=r^2\delta_{r}P_{\mathrm{BS}}/8$, $N_{1,{\mathrm {mode}}}=r\operatorname{Re}(n_{+}) P_{\mathrm{BS}}/16$, and $N_{1,{\mathrm {shot}}}=\delta_{1,{\mathrm {shot}}}$, which represent source-laser intensity noise, reference-laser intensity noise, mode-impurity noise, and shot noise, respectively. In deriving Eq.~\eqref{eq.noisyM1}, we assume that
\begin{equation}
	|\delta_{\mathrm {SL}}|,\ |\delta_r|,\ |\delta_{\phi_r}|,\ |n_{\pm}|,\ |\alpha_{\pm}|\ll 1,
\end{equation}
and retain only terms linear in these small quantities. The reference-state phase fluctuation $\delta_{\phi_r}$ does not contribute at this order. Its leading contribution to the GW-dependent interference term is proportional to $\alpha_{\pm}\delta_{\phi_r}$ and is therefore second order in the perturbative expansion. The remaining noisy outputs $\widetilde{M}_m$, with $m=2,3,4$, can be written in the same form,
\begin{equation}
	\widetilde{M}_m=P_{m,{\mathrm {DC}}}+S_{m,{\mathrm {GW}}}+N_{m,{\mathrm {SL}}}+N_{m,r}+N_{m,{\mathrm {mode}}}	+N_{m,{\mathrm {shot}}},
	\label{eq.noisyMm}
\end{equation}
where $P_{m,{\mathrm {DC}}}=P_{1,{\mathrm {DC}}}$, $N_{m,{\mathrm {SL}}}=N_{1,{\mathrm {SL}}}$, $N_{m,r}=N_{1,r}$, and
\begin{align}
	S_{2,\mathrm {GW}}=&\frac{r\operatorname{Im}(\alpha_+)P_{\mathrm {BS}}}{8},\\
	S_{3,\mathrm {GW}}=&\frac{r\operatorname{Re}(\alpha_-)P_{\mathrm {BS}}}{8},\\
	S_{4,\mathrm {GW}}=&-\frac{r\operatorname{Im}(\alpha_-)P_{\mathrm {BS}}}{8},\\
	N_{2,\mathrm {mode}}=&\frac{r\operatorname{Im}(n_+)P_{\mathrm {BS}}}{16},\\
	N_{3,\mathrm {mode}}=&\frac{r\operatorname{Re}(n_-)P_{\mathrm {BS}}}{16},\\
	N_{4,\mathrm {mode}}=&-\frac{r\operatorname{Im}(n_-)P_{\mathrm {BS}}}{16}.\label{eq.N4mode}
\end{align}

The response function describing the response of the output $\widetilde{M}$ to a small perturbation of a signal or noise variable $s$ is defined as \cite{PhysRevD.95.062003}\begin{equation}
	H_s(f)=\frac {\partial \widetilde{M} (f)}{\partial s (f)}.
\end{equation}
For a given output, the magnitudes of the response functions for the GW signal, source-laser intensity noise, reference-laser intensity noise, and mode-impurity noise are
\begin{align}
	\left|H_{\mathrm {GW}}(f_g)\right|=&\frac{P_{\mathrm {BS}}rNL\gamma k_3}{8\left[1+(f_g/f_p)^2\right]^{1/2}},\\
	\left|H_{\mathrm {SL}}\right|=&\frac{P_{\mathrm {BS}}r^2}{16},\\
	\left|H_r\right|=&\frac{P_{\mathrm {BS}}r^2}{8},\\
	\left|H_{\mathrm {impurity}}\right|=&\frac{P_{\mathrm {BS}}r}{16}.
\end{align}
Here, for outputs that measure a particular quadrature of the complex GW response $h=h_{\mathrm {R}}+ih_{\mathrm {I}}$, $H_{\mathrm {GW}}$ is understood as the response function with respect to the corresponding quadrature. The same convention applies to the complex mode-impurity amplitudes $n_\pm$. Then, if the amplitude spectral density (ASD) of a noise variable $n$ is $A_n(f)$, the corresponding equivalent GW strain sensitivity for a single output is
\begin{equation}
	S_{h,n}^{1/2}(f)=\left|	\frac{H_n}{H_{\mathrm {GW}}(f)}\right|A_n(f).
\end{equation}
Unless otherwise stated, we first evaluate the sensitivity of a single detector output.

Moreover, each optical component in the detector may introduce mode-impurity noise through crosstalk between OAM states. For example, if the phase modulator or mode projectors are implemented using a spatial light modulator (SLM), temporal phase fluctuations of the SLM may lead to imperfect mode generation or projection and thereby introduce unwanted OAM components. Such crosstalk can be reduced by replacing the SLM with static optical elements, such as axicons or geometric-phase elements such as $q$-plates \cite{PhysRevLett.96.163905,PhysRevLett.103.013601}. The optical mode can also be cleaned by the MZI before being injected into the Fabry--Perot cavity. Meanwhile, during $N$ round trips in the cavity, the light undergoes $2N$ reflections from the cavity mirrors. Since $N \gg 1$, we focus on mode-impurity noise accumulated within the cavity. In this work, we consider three dominant mechanisms: angular noise, transverse displacement noise, and mirror-surface imperfections.

\subsection{Laser intensity noise}

We first consider the laser intensity noise contributing to $N_{m,{\mathrm {SL}}}$ and $N_{m,r}$ for the source and reference lasers, respectively. The main contributions to laser intensity noise include thermal noise, generation--recombination noise, and flicker noise ($1/f$ noise) \cite{1449826}. At frequencies below approximately $100~{\mathrm {Hz}}$, the first two contributions can be approximated as white noise \cite{1449826,MARINOV20221}, whereas flicker noise has a power spectral density (PSD) that scales approximately as $1/f$. Correspondingly, its ASD scales as $f^{-1/2}$. Reported laser intensity-noise ASDs can reach approximately $6\times 10^{-5}/\sqrt{\mathrm {Hz}}$ at $0.1~{\mathrm {mHz}}$ \cite{Li:25} and $4\times10^{-9}/\sqrt{\mathrm {Hz}}$ above $30~{\mathrm {Hz}}$ \cite{PhysRevD.111.062002}. To obtain a conservative estimate of the detector sensitivity, we model the intensity-noise ASD as
\begin{equation}
	A_{\mathrm {LP}}(f)=\left[6\times 10^{-7}\left(\frac{f}{1~{\mathrm {Hz}}}\right)^{-1/2}
	+4\times 10^{-9}\right]/\sqrt{\mathrm {Hz}}.
	\label{eq.ALP}
\end{equation}
This model reproduces the reported low-frequency noise level while remaining above the high-frequency value reported in Ref.~\cite{PhysRevD.111.062002}. It therefore provides a conservative estimate of the laser intensity noise over the frequency range considered here. The corresponding strain sensitivities associated with $N_{m,{\mathrm {SL}}}$ and $N_{m,r}$ are
\begin{align}
	S^{1/2}_{h,{\mathrm {SL}}}(f)=&
	\left|\frac{H_{\mathrm {SL}}}{H_{\mathrm {GW}}(f)}\right|A_{\mathrm {LP}}(f)
	\nonumber\\
	=&\left[6\times 10^{-7}\left(\frac{f}{1~{\mathrm {Hz}}}\right)^{-1/2}+4\times10^{-9}\right]\nonumber \\
	&\times 
	\frac{r\left[1+(f/f_p)^2\right]^{1/2}}{2NL\gamma k_3}
	/\sqrt{\mathrm {Hz}},\label{eq.ssl}\\
	S^{1/2}_{h,r}(f)=&\left|\frac{H_r}{H_{\mathrm {GW}}(f)}\right|A_{\mathrm {LP}}(f)=2S^{1/2}_{h,{\mathrm {SL}}}.
	\label{eq.sr}
\end{align}

\subsection{Mode-impurity noise I: Angular noise}

\begin{figure}[tbhp]
	\centering
	\subfloat[][Rotation]{%
		\includegraphics[height=0.26\linewidth]{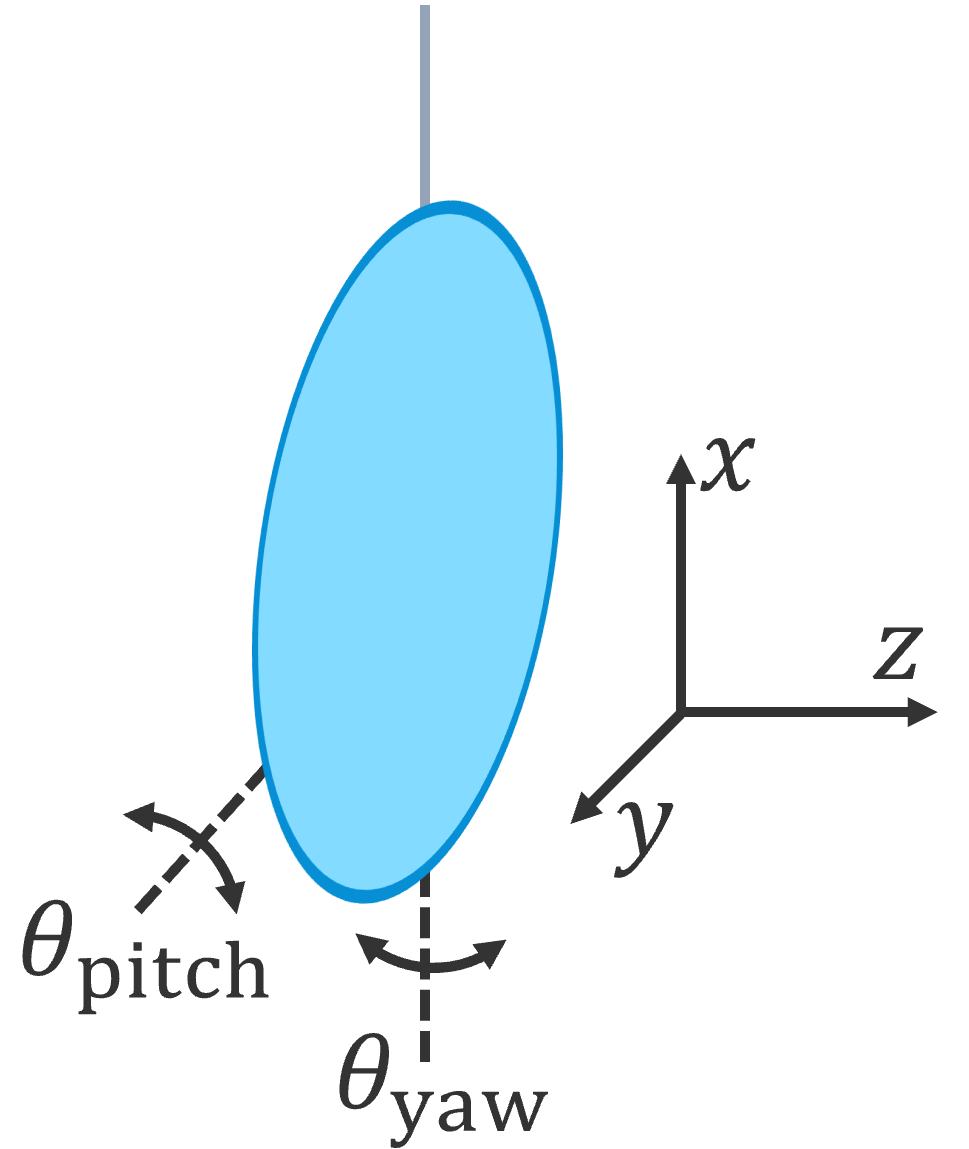}%
		\label{fig.mirror}
	}\hfill
	\subfloat[][Pitch]{%
		\includegraphics[height=0.26\linewidth]{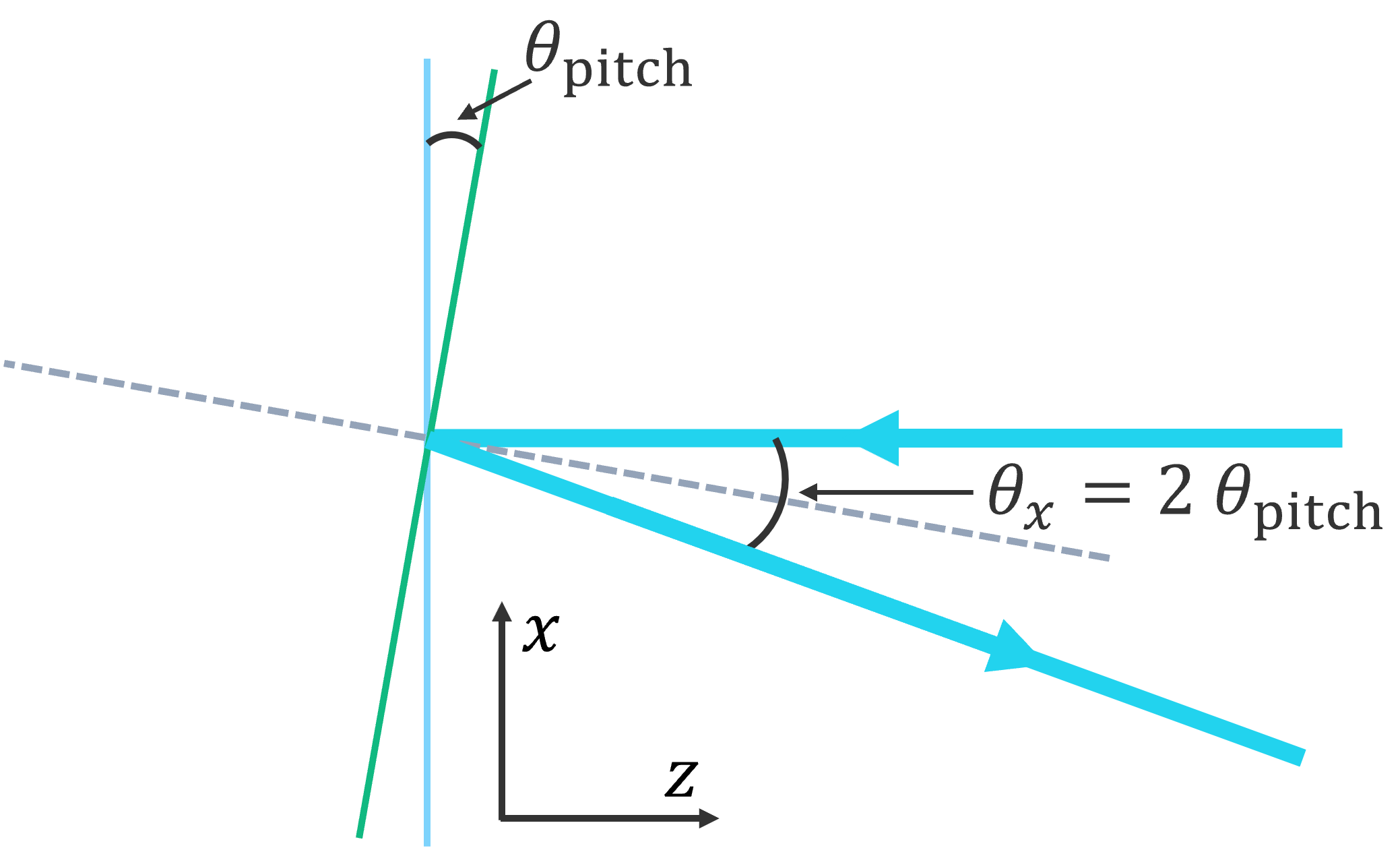}%
		\label{fig.pitch}
	}\hfill
	\subfloat[][Yaw]{%
		\includegraphics[height=0.26\linewidth]{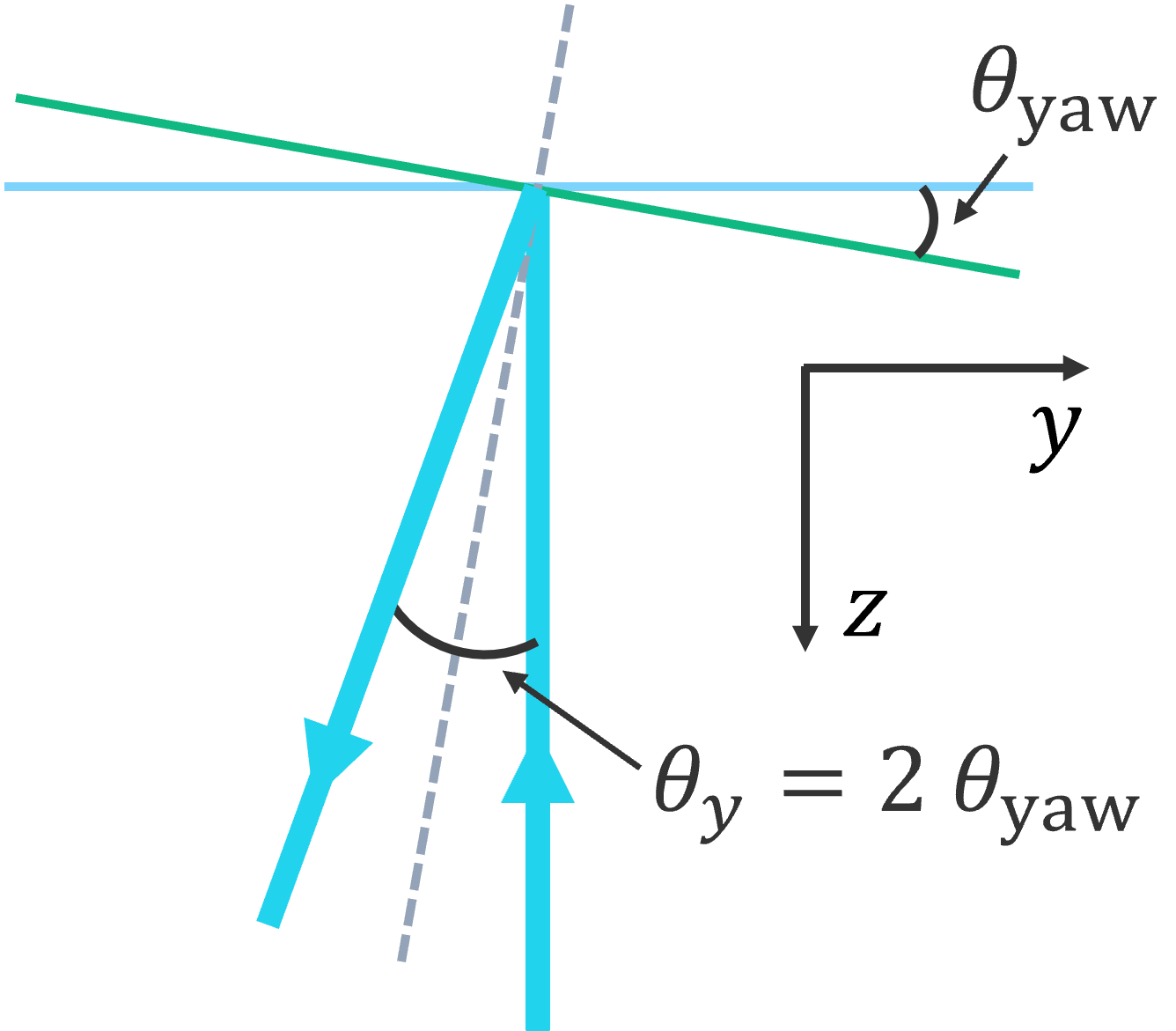}%
		\label{fig.yaw}
	}
	\caption{(a) Mirror rotations that can introduce mode impurity. A beam incident on a mirror can be deflected when the mirror is rotated by a small angle in (b) pitch or (c) yaw. The corresponding angular deflections of the reflected beam are denoted by $\theta_x$ and $\theta_y$, respectively.}
	\label{fig.mirrorRotation}
\end{figure}

A mirror has three rotational degrees of freedom: pitch, yaw, and roll. Figure~\ref{fig.mirrorRotation} illustrates the pitch and yaw rotations of a mirror. Because an ideal mirror is rotationally symmetric about the propagation axis of the incident light, a small roll rotation does not change the field distribution and is therefore neglected. If light is incident on a mirror along the $z$ axis and the mirror is rotated by a small angle $\theta_{\mathrm {pitch}}$ or $\theta_{\mathrm {yaw}}$, the reflected beam is deflected by the angles $\theta_x=2\theta_{\mathrm {pitch}}$ or $\theta_y=2\theta_{\mathrm {yaw}}$, respectively, as shown in Figs.~\ref{fig.pitch} and \ref{fig.yaw}.

\begin{figure}[tbhp]
	\centering
	\includegraphics[width=0.6\linewidth]{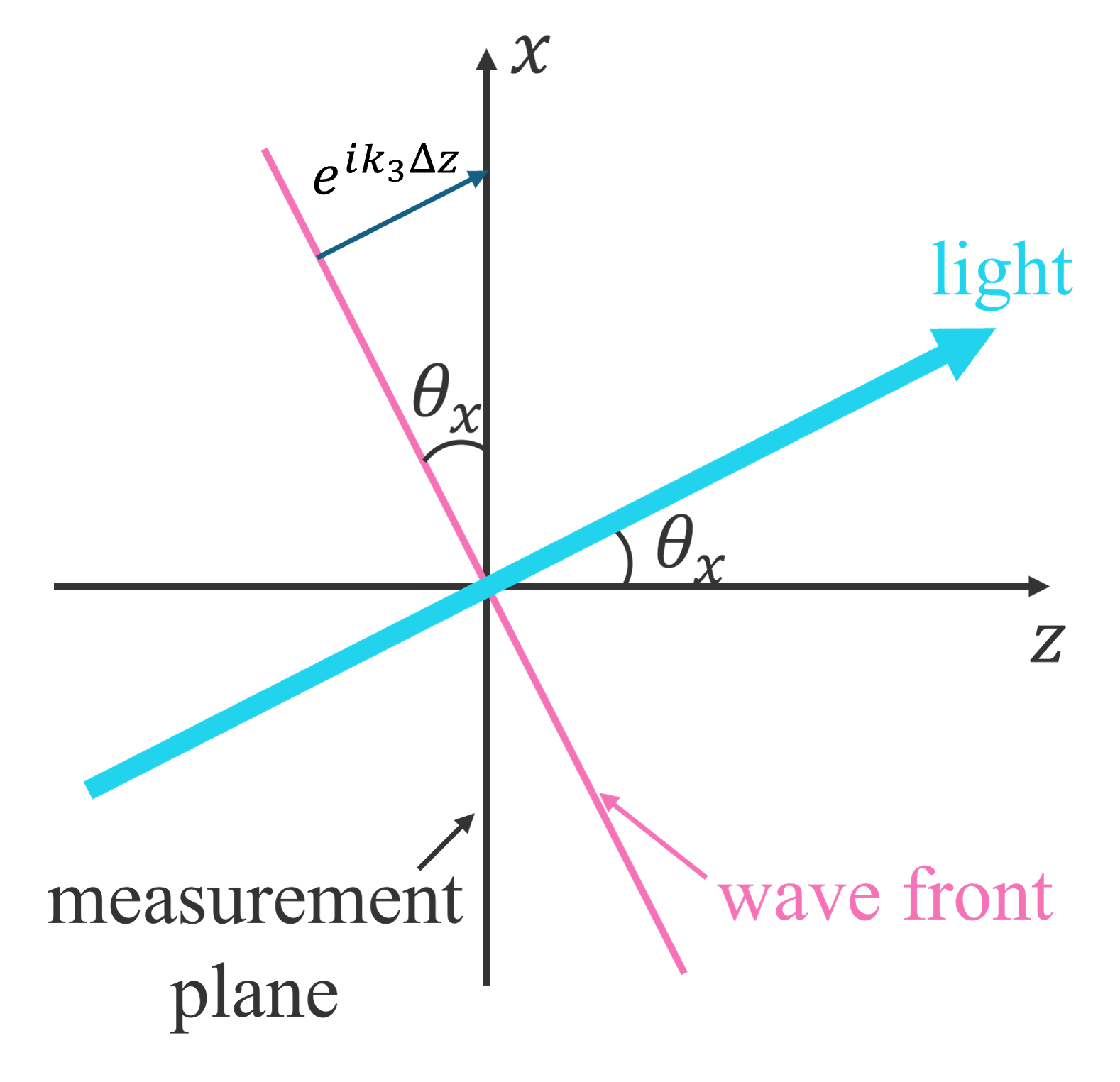}
	\caption{Phase variation induced by a small angular deflection $\theta_x$ of the reflected beam. Relative to the original transverse measurement plane, the optical path length varies by $\Delta z=x\sin\theta_x\approx x\theta_x$, producing the phase factor $e^{ik_3\Delta z}$.}
	\label{fig.measurePlane}
\end{figure}

Suppose the light field is measured on a plane transverse to its original propagation direction. For a beam propagating along the $z$ axis, this measurement plane coincides with the $xy$ plane. If the beam is deflected by a small angle $\theta_x$ owing to a pitch rotation of the mirror, the optical path length across the measurement plane acquires a variation $\Delta z=x\sin\theta_x\approx x\theta_x$, as illustrated in Fig.~\ref{fig.measurePlane}. The corresponding field therefore acquires an additional phase factor $e^{ik_3\Delta z}\approx e^{ik_3x\theta_x}$. Similarly, a yaw rotation produces $\Delta z\approx y\theta_y$ and an additional phase factor $e^{ik_3y\theta_y}$. Suppose that the incident light field is initially prepared in the OAM state $\ket{l}$. After reflection from the tilted mirror, the angular perturbation modifies the state as $ e^{ik_3(x\theta_x+y\theta_y)}\ket{l}$. To first order in $\theta_x$ and $\theta_y$, this state becomes
\begin{align}
	&\left(1+ik_3x\theta_x+ik_3y\theta_y\right)\ket{l}\nonumber\\
	=&\left(1+ik_3\theta_x\rho\cos\phi+ik_3\theta_y\rho\sin\phi\right)\ket{l}\nonumber\\
	=&\left(1+c_+\rho e^{i\phi}+c_-\rho e^{-i\phi}\right)\ket{l}\nonumber\\
	=&\ket{l}+\tilde n_{l,+}\ket{l+1}+\tilde n_{l,-}\ket{l-1},\label{eq.stateDueToAngleNoise}
\end{align}
where $c_\pm=ik_3 (\theta_x\mp i\theta_y )/2$, and $\tilde n_{l,\pm}=\bra{l\pm1}c_\pm\rho e^{\pm i\phi}\ket{l}$ denotes the angular-noise-induced coupling amplitude from the initial OAM state $\ket{l}$ to the neighboring state $\ket{l\pm1}$.

The OAM state $\ket{l}$ can be expanded in the transverse position basis as $\braket{\mathbf{x}_\perp}{l}=N_l J_l(k_\perp\rho)e^{il\phi}$, where $N_l$ is the normalization factor. We normalize the state over a circular transverse region of radius $w$, such that $\braket{l}{l}=\int d^2\mathbf{x}_\perp N_l^2J_l^2(k_\perp\rho)=2\pi N_l^2 \int_0^w \rho d\rho	J_l^2(k_\perp\rho)=1$. Using \begin{equation}
\int_0^w \rho d\rho J_l^2(k_\perp\rho)=\frac{w^2}{2}
	\left[J_l^2(k_\perp w)-J_{l-1}(k_\perp w)J_{l+1}(k_\perp w)\right],
\end{equation}
the normalization factor is
\begin{equation}
	N_l	= \left (\pi w^2\left[J_l^2(k_\perp w)-J_{l-1}(k_\perp w)J_{l+1}(k_\perp w)	\right]
	\right )^{-1/2}.
\end{equation}
The angular-noise-induced coupling coefficients $\tilde n_{l,\pm}$ are then
\begin{align}
	\tilde n_{l,\pm}=&\bra{l\pm1}c_\pm\rho e^{\pm i\phi}\ket{l}\nonumber\\
	=&\int d^2\mathbf{x}_\perp \braket{l\pm1}{\mathbf{x}_\perp}	c_\pm\rho e^{\pm i\phi}
	\braket{\mathbf{x}_\perp}{l}\nonumber\\
	=&2\pi c_\pm N_{l\pm1}N_l\int_0^w\rho^2 d\rho J_l(k_\perp\rho)J_{l\pm1}(k_\perp\rho)
	\nonumber\\
	=&c_\pm k_\perp^{-1}\Big (\pm J_{l\pm1}^2(k_\perp w)+l [J_{l\pm1}^2(k_\perp w)
		-J_l(k_\perp w)\nonumber \\ &\times J_{l\pm2}(k_\perp w) ]\Big ) \Big( 
			 [J_{l\pm1}^2(k_\perp w)-J_l(k_\perp w)J_{l\pm2}(k_\perp w) ]\nonumber \\ &\times
			[J_l^2(k_\perp w)-J_{l-1}(k_\perp w)J_{l+1}(k_\perp w)]	\Big )^{-1/2}.
\end{align}
In deriving the last expression, we use the identity
\begin{align}
	&\rho^2 J_l(k_\perp\rho)J_{l\pm1}(k_\perp\rho)\nonumber \\
	=&\pm\frac{1}{k_\perp}\frac{d}{d\rho}\left[
	\frac{\rho^2}{2}J_{l\pm1}^2(k_\perp\rho)\right]+
	\frac{l\rho}{k_\perp}J_{l\pm1}^2(k_\perp\rho).
\end{align}
For fixed $l$ and $k_\perp w\gg1$, the Bessel functions can be approximated by $J_l(k_\perp w)\approx [2/(\pi k_\perp w)]^{1/2}\cos\left(k_\perp w-l\pi/2-\pi/4 \right)$. The coupling coefficients then reduce to
\begin{equation}
	\tilde n_{l,\pm}\approx\frac{c_\pm}{k_\perp}\left[
	l\pm\sin^2\left(k_\perp w-\frac{l\pi}{2}-\frac{\pi}{4}\right)\right].
\end{equation}

The coupling coefficients can be further simplified by allowing the measurement radius $w$ to fluctuate slightly around its mean value $\bar w$, such that $w=\bar w+\delta w$. We assume that $\delta w$ follows a Gaussian distribution, $P(\delta w)= \exp[-\delta w^2/(2\sigma_w^2)]/(\sqrt{2\pi}\sigma_w)$, where $\sigma_w$ is the standard deviation. The ensemble average of the oscillatory factor is then
\begin{align}
	&\left\langle\sin^2\left(k_\perp w-\frac{l\pi}{2}-\frac{\pi}{4}\right)\right\rangle	\nonumber\\
	=&\frac{1}{2}-\frac{1}{2}(-1)^l\left\langle\sin\left(2k_\perp\bar w+2k_\perp\delta w	\right)\right\rangle\nonumber\\
	=&\frac 1 2 -\frac 1 2 (-1)^l \left\langle [\sin(2k_\perp \bar w)\cos(2k_\perp \delta w)\right .\nonumber \\
	&\left .+\cos(2k_\perp \bar w)\sin(2k_\perp \delta w)]\right\rangle\nonumber \\
	=&\frac{1}{2}-\frac{1}{2}(-1)^l\sin(2k_\perp\bar w)\left\langle	\cos(2k_\perp\delta w)	\right\rangle\nonumber\\
	=&\frac{1}{2}-\frac{1}{2}(-1)^l\sin(2k_\perp\bar w)\exp\left(-2k_\perp^2\sigma_w^2\right).
\end{align}
Here, the term proportional to $\langle\sin(2k_\perp\delta w)\rangle$ vanishes because the probability distribution $P(\delta w)$ is symmetric about $\delta w=0$. When the fluctuations satisfy $k_\perp^{-1}\ll\sigma_w\ll\bar w$, the exponential factor is strongly suppressed, and hence $\left\langle\sin^2\left(k_\perp w-l\pi/2-\pi/4 \right)\right\rangle\approx 1/2$. The angular-noise-induced coupling coefficients therefore reduce to
\begin{equation}
	\tilde n_{l,\pm}\approx
	\frac{1}{\gamma}\left(\pm\theta_{\mathrm {yaw}}+i\theta_{\mathrm {pitch}}\right)\left(l\pm\frac{1}{2}	\right),
\end{equation}
where we have used $\gamma=k_\perp/k_3$, $\theta_x=2\theta_{\mathrm {pitch}}$, and $\theta_y=2\theta_{\mathrm {yaw}}$.

Since the light undergoes $2N$ reflections from the cavity mirrors, the angular-noise-induced mode crosstalk accumulates over the round trips. If the angular fluctuations are fully coherent over the cavity storage time, the corresponding ASD scales linearly with the number of reflections, giving a factor of $2N$. In contrast, if the fluctuations at different reflections are statistically independent, the ASD accumulates incoherently and scales as $\sqrt{2N}$. To obtain a conservative estimate of the angular-noise contribution, we adopt the coherent scaling and write $n_{l,\pm}=2N\tilde n_{l,\pm}$. The detector does not measure $n_{l,\pm}$ individually. Instead, as shown in Eqs.~\eqref{eq.noisyM1}--\eqref{eq.N4mode}, the relevant combinations are
\begin{align}
	n_+	=n_{l,+}+n_{-l,+}=&\frac{2N}{\gamma}\left(\theta_{\mathrm {yaw}}+i\theta_{\mathrm {pitch}}\right),
	\label{eq.n+}\\
	n_-	=n_{l,-}+n_{-l,-}=&\frac{2N}{\gamma}\left(\theta_{\mathrm {yaw}}-i\theta_{\mathrm {pitch}}\right).	\label{eq.n-}
\end{align}
The two coefficients therefore satisfy
\begin{equation}
	n_+=n_-^*.\label{eq.noiseCoefficientRelation}
\end{equation}

Reference~\cite{Martinez-Rincon:17} reports the ASD of angular noise for an experimental system mounted on top of a large concrete pillar resting on sand and mechanically isolated from the surrounding building. Based on Fig.~4 of Ref.~\cite{Martinez-Rincon:17}, we approximate the angular-noise ASD as
\begin{align}
	A_\theta(f)=&
	\Bigg[2\times10^{-13}+\frac{5\times10^{-14}}{f/(1~{\mathrm {Hz}})}+\frac{5\times10^{-11}}
	{\left[f/(0.16~{\mathrm {Hz}})\right]^4+1}\Bigg]\nonumber\\
	&\times {\mathrm {rad}}/\sqrt{\mathrm {Hz}}.
\end{align}
Using Eqs.~\eqref{eq.n+} and \eqref{eq.n-}, the magnitude of the response function for angular noise is
\begin{equation}
	\left|H_\theta\right|=\left|H_{\mathrm {impurity}}\right|\frac{2N}{\gamma}
	=\frac{NP_{\mathrm {BS}}r}{8\gamma}.
\end{equation}
The corresponding equivalent GW strain sensitivity is therefore
\begin{align}
	S^{1/2}_{h,\theta}(f)=&\left|	\frac{H_\theta}{H_{\mathrm {GW}}(f)}\right|A_\theta(f)\nonumber \\	
	=&\frac{\left[1+(f/f_p)^2\right]^{1/2}}{\gamma^2 k_3L}
	\Bigg[2\times10^{-13}+\frac{5\times10^{-14}}{f/(1~{\mathrm {Hz}})}\nonumber\\
	&+\frac{5\times10^{-11}}{\left[f/(0.16~{\mathrm {Hz}})\right]^4+1}\Bigg]
	/\sqrt{\mathrm {Hz}}.\label{eq.stheta}
\end{align}

\subsection{Mode-impurity noise II: Transverse displacement noise}

\begin{figure}[tbhp]
	\centering
	\includegraphics[width=0.6\linewidth]{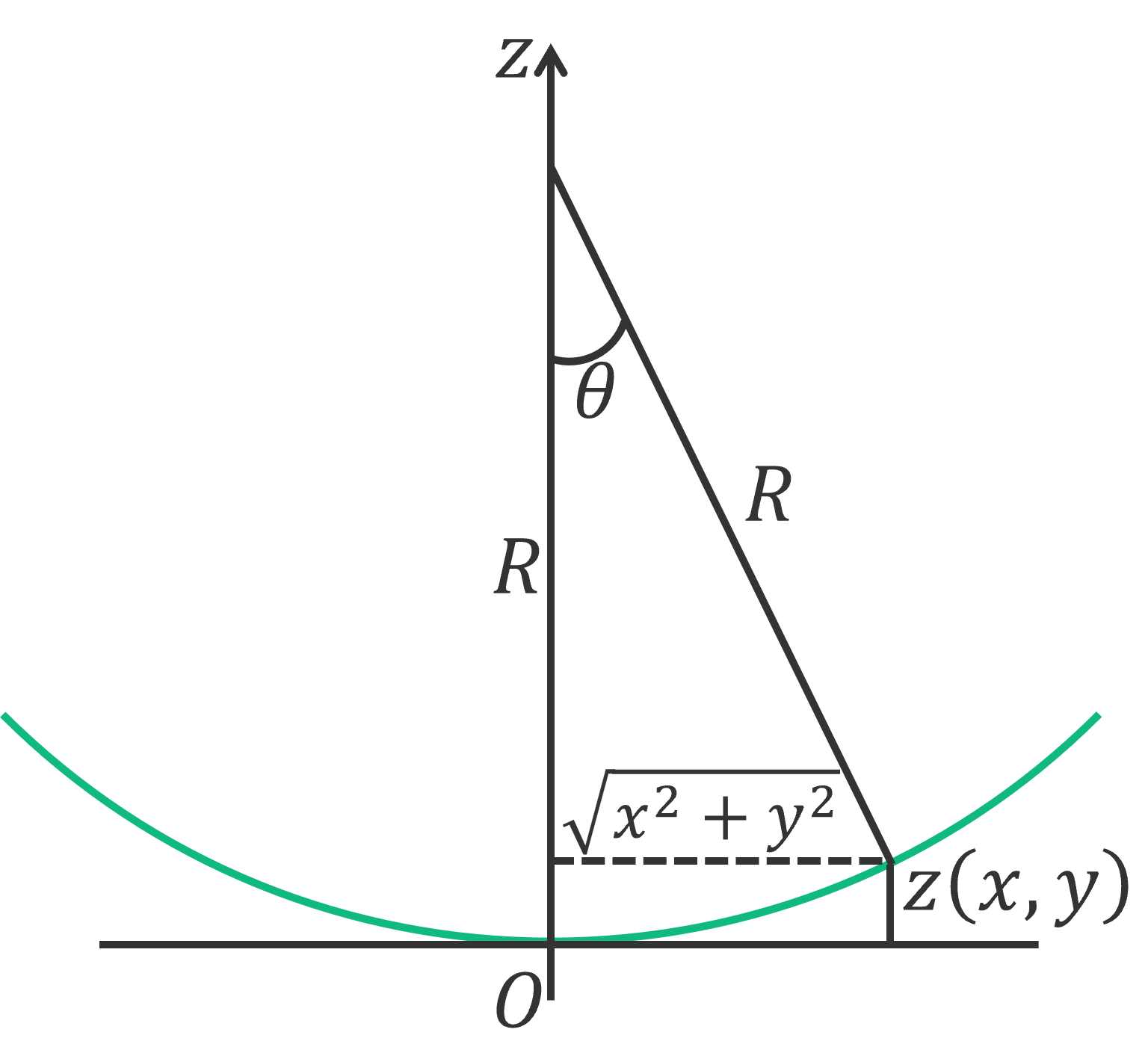}
	\caption{Schematic of the curved wavefront of a Bessel--Gaussian beam. A transverse displacement of the beam relative to the mirror changes the optical path length across the transverse plane and thereby introduces an additional spatially dependent phase.}
	\label{fig.sphericalFront}
\end{figure}

For an ideal Bessel mode, the longitudinal phase is proportional to $e^{ik_3z}$, so its wavefront is planar and a transverse displacement of the beam on an ideal mirror does not change the light field. In practice, however, only Bessel--Gaussian modes can be generated. The wavefront of a Bessel--Gaussian beam inherits the curvature of the Gaussian envelope, with a radius of curvature $R(L)=L\left(1+z_R^2/L^2\right)$ at a distance $L$ from the beam waist. As illustrated in Fig.~\ref{fig.sphericalFront}, in the absence of transverse displacement, the longitudinal separation between the spherical wavefront and the transverse plane is $z(x,y)=	R-(R^2-x^2-y^2)^{1/2}\approx (x^2+y^2)/(2R)$, where the paraxial approximation $x^2+y^2\ll R^2$ has been used. If the beam is displaced transversely by small amounts $\delta x$ and $\delta y$, this separation becomes $z'(x,y)\approx [(x-\delta x)^2+(y-\delta y)^2]/(2R)$. Neglecting terms of second order in the displacements, the resulting change in optical path length is
\begin{equation}
	\Delta z(x,y)=z'(x,y)-z(x,y)\approx	-\frac{x\delta x+y\delta y}{R}.
\end{equation}
An initial OAM state $\ket{l}$ therefore acquires an additional phase,
\begin{equation}
	e^{ik_3\Delta z(x,y)}\ket{l}\approx\left(
	1-\frac{ik_3x\delta x}{R}-\frac{ik_3y\delta y}{R}\right)\ket{l}.
\end{equation}
Comparing this expression with the first line of Eq.~\eqref{eq.stateDueToAngleNoise}, we identify the equivalent mirror-angle fluctuations as
\begin{equation}
	\theta_{\mathrm {pitch}}=-\frac{\delta x}{2R},\qquad
	\theta_{\mathrm {yaw}}=-\frac{\delta y}{2R}.
\end{equation}
Using Eqs.~\eqref{eq.n+} and \eqref{eq.n-}, the corresponding mode-crosstalk coefficients are therefore
\begin{align}
	n_+	=&-\frac{N}{R\gamma}\left(\delta y+i\delta x\right),\\
	n_-	=&-\frac{N}{R\gamma}\left(\delta y-i\delta x\right).
\end{align}
These coefficients also satisfy the relation in Eq.~\eqref{eq.noiseCoefficientRelation}, $n_+=n_-^*$. Following the same conservative assumption used for angular noise, we take the transverse-displacement-induced crosstalk to accumulate coherently over the $2N$ mirror reflections. The magnitude of the corresponding response function is then
\begin{equation}
	\left|H_{\mathrm {D}}\right|=	\left|H_{\mathrm {impurity}}\right|\frac{N}{R\gamma}=
	\frac{NP_{\mathrm {BS}}r}{16R\gamma}.
\end{equation}

The displacement-noise level is strongly site dependent. Motivated by the conversion from acceleration noise to displacement noise at low frequencies \cite{bormann1998conversion}, as well as the approximate $f^{-2}$ scaling of the displacement ASD associated with low-frequency gravity-gradient, seismic, and radiation-pressure disturbances \cite{PhysRevD.93.112004}, we model the displacement-noise ASD of an isolated platform as
\begin{equation}
	A_{\mathrm {D}}(f)=\frac{10^{-11}}{\left[f/(1~{\mathrm {Hz}})\right]^2} {\mathrm {m}}/\sqrt{\mathrm {Hz}}.
	\label{eq.AD}
\end{equation}
At $f=1~{\mathrm {mHz}}$ and $1~{\mathrm {Hz}}$, this model gives $A_{\mathrm {D}}=10^{-5}~{\mathrm {m}}/\sqrt{\mathrm {Hz}}$ and $10^{-11}~{\mathrm {m}}/\sqrt{\mathrm {Hz}}$, respectively, consistent with the displacement-noise levels reported in Refs.~\cite{bormann1998conversion,MATICHARD2015287}. At $f=10~{\mathrm {Hz}}$ and $100~{\mathrm {Hz}}$, it gives $A_{\mathrm {D}}=10^{-13}~{\mathrm {m}}/\sqrt{\mathrm {Hz}}$ and $10^{-15}~{\mathrm {m}}/\sqrt{\mathrm {Hz}}$, respectively, which are substantially higher than the corresponding displacement-noise levels achieved in LIGO. We therefore use Eq.~\eqref{eq.AD} as a conservative upper-envelope model for the displacement noise over the frequency range considered here. Using the response function derived above, the corresponding equivalent GW strain sensitivity is
\begin{equation}
	S^{1/2}_{h,\mathrm {D}}(f)=\frac{\left[1+(f/f_p)^2\right]^{1/2}}{2R\gamma^2k_3L}
	\frac{10^{-11}}{\left[f/(1~{\mathrm {Hz}})\right]^2}{\mathrm {m}}/\sqrt{\mathrm {Hz}}.
	\label{eq.sd}
\end{equation}

\subsection{Mode-impurity noise III: Mirror surface imperfection noise}

\begin{figure}[tbhp]
	\centering
	\includegraphics[width=0.6\linewidth]{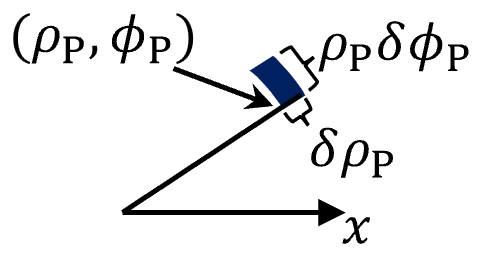}
	\caption{Schematic of a localized mirror-surface imperfection centered at $(\rho_{\mathrm {P}},\phi_{\mathrm {P}})$. Its radial and azimuthal extents are $\delta\rho_{\mathrm {P}}$ and $\rho_{\mathrm {P}}\delta\phi_{\mathrm {P}}$, respectively.}
	\label{fig.imperfection}
\end{figure}

Surface imperfections can be introduced during substrate fabrication, dielectric-coating deposition, handling, and cleaning. Such imperfections may perturb the reflected light field and thereby induce crosstalk between OAM states. We consider a localized surface imperfection centered at $(\rho_{\mathrm {P}},\phi_{\mathrm {P}})$ on the transverse plane, as illustrated in Fig.~\ref{fig.imperfection}. The imperfection has a radial extent $\delta\rho_{\mathrm {P}}$, an azimuthal extent $\rho_{\mathrm {P}}\delta\phi_{\mathrm {P}}$, and a height $h_{\mathrm {P}}$ relative to the nominal mirror surface.

Because a surface imperfection locally changes the optical path length, reflection from an imperfection of height $h_{\mathrm {P}}$ introduces an additional phase. The state $\ket{l}$ therefore transforms as
\begin{equation}
	e^{-2ik_3h_{\mathrm {P}}\Theta_{\mathrm {P}}(\rho,\phi)}\ket{l}
	\approx\left[1-2ik_3h_{\mathrm {P}}\Theta_{\mathrm {P}}(\rho,\phi)\right]\ket{l},
\end{equation}
where $\Theta_{\mathrm {P}}(\rho,\phi)=\Theta(\rho-\rho_{\mathrm {P}})\Theta(\rho_{\mathrm {P}}+\delta\rho_{\mathrm {P}}-\rho)\Theta(\phi-\phi_{\mathrm {P}})\Theta(\phi_{\mathrm {P}}+\delta\phi_{\mathrm {P}}-\phi)$, and $\Theta(\cdot)$ denotes the Heaviside step function. The function $\Theta_{\mathrm {P}}(\rho,\phi)$ restricts the additional phase to the area occupied by the surface imperfection. Because the detector is sensitive to crosstalk into the neighboring OAM state $\ket{l\pm1}$, we retain only these components and write $\left[	1-2ik_3h_{\mathrm {P}}\Theta_{\mathrm {P}}(\rho,\phi)\right]\ket{l}	\approx	\ket{l}	+n_{l,+}\ket{l+1}+n_{l,-}\ket{l-1}$, where $n_{l,\pm}=\bra{l\pm1}\left[
-2ik_3h_{\mathrm {P}}\Theta_{\mathrm {P}}(\rho,\phi)\right]\ket{l}$. For $k_\perp w\gg1$, $\delta\rho_{\mathrm {P}}\ll 1\mathrm{~m}$, and $\delta\phi_{\mathrm {P}}\ll 1$, the coupling coefficients can be approximated as
\begin{equation}
	n_{l,\pm}=-\frac{ik_3k_\perp h_{\mathrm {P}}}{w}
	J_{l\pm1}(k_\perp\rho_{\mathrm {P}})J_l(k_\perp\rho_{\mathrm {P}})
	e^{\mp i\phi_{\mathrm {P}}}\delta A,
\end{equation}
where $\delta A=\rho_{\mathrm {P}}\delta\rho_{\mathrm {P}}\delta\phi_{\mathrm {P}}$ is the area of the imperfection. For simplicity, we assume $\delta A\approx (\delta\rho_{\mathrm {P}})^2$. The combinations relevant to the detector outputs are then
\begin{align}
	n_\pm=&	n_{l,\pm}+n_{-l,\pm}\nonumber\\
	=&\pm\frac{2ik_3k_\perp h_{\mathrm {P}}\delta A}{w}e^{\mp i\phi_{\mathrm {P}}}
	J_l(k_\perp\rho_{\mathrm {P}})J_l^{(1)}(k_\perp\rho_{\mathrm {P}}),
\end{align}
where $J_l^{(1)}(x)=dJ_l(x)/dx$. These coefficients satisfy $n_+=n_-^*$, consistent with Eq.~\eqref{eq.noiseCoefficientRelation}. The magnitude of the response function from the surface-height perturbation $h_{\mathrm {P}}$ to the detector output is therefore
\begin{align}
	\left|H_{\mathrm {P}}\right|=&\left|H_{\mathrm {impurity}}\right|
	\frac{2k_3k_\perp\delta A	\left|
		J_l(k_\perp\rho_{\mathrm {P}})J_l^{(1)}(k_\perp\rho_{\mathrm {P}})\right|}{w}
	\nonumber\\
	=&\frac{P_{\mathrm {BS}}r k_3k_\perp\delta A
		\left|	J_l(k_\perp\rho_{\mathrm {P}})J_l^{(1)}(k_\perp\rho_{\mathrm {P}})\right|
	}{8w}.
\end{align}

The surface imperfection itself is static and therefore produces only static mode crosstalk. Under high-power illumination, however, a localized absorbing region can undergo thermal deformation. We model this effect using the point-absorber model of Ref.~\cite{Brooks:21}. Within the absorbing region, the magnitude of the surface deformation can be approximated as
\begin{equation}
	\Delta h_{\mathrm {P}}(r_d)\approx\frac{\alpha_{\mathrm{th}} P_{\mathrm {abs}}}{4\pi\kappa}
	\left(\frac{r_d}{\delta\rho_{\mathrm {P}}}\right)^2,\qquad r_d\leq\delta\rho_{\mathrm {P}},
\end{equation}
where $r_d$ is the distance from the center of the absorber, $\alpha_{\mathrm{th}}$ is the thermal expansion coefficient, and $\kappa$ is the thermal conductivity. Here, $\delta\rho_{\mathrm {P}}$ is identified with the characteristic radius of the absorbing region. For simplicity, we use the maximum deformation within this region, obtained at $r_d=\delta\rho_{\mathrm {P}}$,
\begin{equation}
	\Delta h_{\mathrm {P}}\approx\frac{\alpha_{\mathrm{th}} P_{\mathrm {abs}}}{4\pi\kappa}.
\end{equation}
Following Ref.~\cite{Brooks:21}, the absorbed power $P_{\mathrm {abs}}$ is approximated by the local optical intensity multiplied by the effective absorbing area,
\begin{equation}
	P_{\mathrm {abs}}=\delta A \frac{NP_{\mathrm {BS}}}{\pi w_m^2}
	\exp\left[	-\left(	\frac{\rho_{\mathrm {P}}}{w_m}\right)^2\right],
\end{equation}
where $w_m$ denotes the optical beam radius on the mirror. The factor $N$ in $P_{\mathrm {abs}}$ accounts for the accumulated optical power over the $N$ cavity round trips. Therefore, no additional factor of $N$ is included in the mode-crosstalk response function $H_{\mathrm {P}}$, which would otherwise double count the cavity enhancement. This expression treats $\delta A$ as an effective absorbing area, consistent with the point-absorber approximation of Ref.~\cite{Brooks:21}. Fluctuations in the incident laser intensity induce corresponding fluctuations in the absorbed power and hence in the thermal deformation. Neglecting the finite thermal-response time, the ASD of the surface-height fluctuation is
\begin{equation}
	A_{\mathrm {P}}(f)=\frac{\alpha_{\mathrm{th}} P_{\mathrm {abs}}}{4\pi\kappa}A_{\mathrm {LP}}(f),
\end{equation}
where $A_{\mathrm {LP}}(f)$ is the relative laser-intensity-noise ASD given in Eq.~\eqref{eq.ALP}. The corresponding equivalent GW strain sensitivity is therefore
\begin{align}
	S^{1/2}_{h,\mathrm {P}}(f)=&\left|\frac{H_{\mathrm {P}}}{H_{\mathrm {GW}}(f)}\right|A_{\mathrm {P}}(f)
	\nonumber\\
	=&\frac{P_{\mathrm {BS}}k_3\alpha_{\mathrm{th}}(\delta A)^2	\left|
		J_l(k_\perp\rho_{\mathrm {P}})J_l^{(1)}(k_\perp\rho_{\mathrm {P}})	\right|e^{-(
			\rho_{\mathrm {P}}/w_m )^2}}{	4\pi^2 w w_m^2\kappa L}\nonumber\\
	&\times	\left[1+\left(\frac{f}{f_p}\right)^2\right]^{1/2}\Bigg [
	6\times10^{-7}\left(\frac{f}{1~{\mathrm {Hz}}}\right)^{-1/2}\nonumber \\
	&+4\times 10^{-9}\Bigg ] 1/\sqrt{\mathrm {Hz}}.
	\label{eq.sp}
\end{align}
Following the representative parameters considered in Ref.~\cite{Brooks:21}, we take $\alpha_{\mathrm{th}}=0.55\times 10^{-6}~\mathrm{K}^{-1}$, $\kappa=1.38~ \mathrm{W}\mathrm{m}^{-1}\mathrm{K}^{-1}$, $\rho_{\mathrm {P}}=20~{\mathrm {mm}}$ and $\delta\rho_{\mathrm {P}}=40~\mu{\mathrm{m}}$. We further assume $w=w_m=0.4~{\mathrm {m}}$ and $l=2$.

\subsection{Shot noise}

For each detector output, let $\bar P$ denote the time-averaged optical power measured over an observation time $T$. The corresponding mean photon number is $\bar N_\gamma=\bar P T/\omega$, where $\omega$ is the photon frequency and we use $\hbar=1$. For a coherent light field, the detected photon number follows Poisson statistics, such that
\begin{equation}
	\sigma_\gamma^2=\bar N_\gamma.
\end{equation}
The corresponding root-mean-square fluctuation of the time-averaged optical power is therefore
\begin{equation}
	P_{\mathrm {shot}}=\frac{\sigma_\gamma\omega}{T}=\left(\frac{\bar P\omega}{T}\right)^{1/2}.
\end{equation}
Since the detected power is dominated by the DC reference field, we take $\bar P\simeq P_{m,\mathrm {DC}}=r^2P_{\mathrm {BS}}/16$. The shot-noise fluctuation is then
\begin{equation}
	P_{\mathrm {shot}}=\left(	\frac{r^2P_{\mathrm {BS}}\omega}{16T}	\right)^{1/2}.
\end{equation}
Using the one-sided power spectral density convention, for which the PSD of shot noise is given by $S_{\mathrm {shot}}=2T P_{\mathrm {shot}}^2$, the corresponding ASD is
\begin{equation}
	A_{\mathrm {shot}}=S_{\mathrm {shot}}^{1/2}=\left(\frac{r^2P_{\mathrm {BS}}\omega}{8}\right)^{1/2}.
\end{equation}
The equivalent GW strain sensitivity due to shot noise is therefore
\begin{align}
	S^{1/2}_{h,{\mathrm {shot}}}(f)=&\frac{A_{\mathrm {shot}}}{\left|H_{\mathrm {GW}}(f)\right|}\nonumber \\
	=&\frac{\left[1+(f/f_p)^2\right]^{1/2}}
	{NL\gamma}\left(\frac{8}{k_3 P_{\mathrm {BS}}}\right)^{1/2},\label{eq.sshot}
\end{align}
where, for the paraxial modes considered here, $\omega=(k_3^2+k_\perp^2)^{1/2}\simeq k_3$.

\subsection{Total sensitivity}

In the preceding subsections, we have evaluated the strain sensitivities associated with different noise sources for a single detector output. Because the four outputs have identical noise sensitivities, we now consider the sensitivity obtained by jointly using all four outputs. The shot noise generated at each photodetector is statistically independent of that in the other detectors. However, since the transition amplitudes satisfy $\alpha_+=-\alpha^*_-$, the GW-induced signal contributions satisfy $S_{1,\mathrm{GW}}=-S_{3,\mathrm{GW}}$ for outputs $M_1$ and $M_3$, and $S_{2,\mathrm{GW}}=-S_{4,\mathrm{GW}}$ for outputs $M_2$ and $M_4$. Therefore, the four outputs are combined pairwise to measure the real and imaginary quadratures of the effective GW response, respectively. In each pair, the GW-induced signal contributions add coherently, whereas the independent shot-noise contributions add in quadrature. The resulting sensitivity for each quadrature is improved by a factor of $\sqrt{2}$. Combining the two independent quadratures, the total shot-noise-limited strain sensitivity obtained using all four outputs is
\begin{equation}
	\frac{1}{\sqrt{2}}S^{1/2}_{h,\mathrm {shot}}(f).
\end{equation}

The remaining noise sources have a different statistical property. As shown in Fig.~\ref{fig.LightPath}, the superposition of the signal field $\ket{\psi}$ and the reference field $r\ket{l'}$ is divided into four outputs by the BSs and subsequently measured by different mode projectors and photodetectors. Fluctuations already present in $\ket{\psi}$ or $r\ket{l'}$ before this splitting are therefore inherited by all four outputs and are correlated among them. Consequently, jointly using the four outputs does not provide the $1/\sqrt{2}$ reduction obtained for independent shot noise. Under the assumption of fully correlated fluctuations, their equivalent strain sensitivities remain equal to those obtained for a single output. The total strain sensitivity obtained from the four outputs is therefore
\begin{align}
	S^{1/2}_{h,\epsilon=2}=&
	\Bigg[\left(S^{1/2}_{h,\mathrm {SL}}\right)^2
	+\left(S^{1/2}_{h,r}\right)^2
	+\left(S^{1/2}_{h,\theta}\right)^2
	+\left(S^{1/2}_{h,\mathrm {D}}\right)^2
	\nonumber\\
	&+\left(S^{1/2}_{h,\mathrm {P}}\right)^2
	+\frac{1}{2}\left(S^{1/2}_{h,\mathrm {shot}}\right)^2\Bigg]^{1/2}.\label{eq.noBalancedSensitivity}
\end{align}
Here, $S^{1/2}_{h,\mathrm {SL}}$, $S^{1/2}_{h,r}$, $S^{1/2}_{h,\theta}$, $S^{1/2}_{h,\mathrm {D}}$, $S^{1/2}_{h,\mathrm {P}}$, and $S^{1/2}_{h,\mathrm {shot}}$ are given by Eqs.~\eqref{eq.ssl}, \eqref{eq.sr}, \eqref{eq.stheta}, \eqref{eq.sd}, \eqref{eq.sp}, and \eqref{eq.sshot}, respectively. The parameter $\epsilon$ characterizes the imbalance of the balanced-detection scheme introduced in the next section. The case $\epsilon=2$ corresponds to operation without balanced detection, as defined below.

\section{Balanced detection}

In the preceding section, we derived the total sensitivity obtained by jointly using the four detector outputs. This sensitivity can be further improved using balanced detection \cite{kumar2012versatile,Jia_2023,Bruynsteen:21,Jin:15,8444971}. To illustrate the principle, we first compare the noisy outputs $\widetilde M_1$ and $\widetilde M_3$. Using the notation introduced above, $\widetilde M_3$ can be written as
\begin{align}
	\widetilde M_3=&P_{3,\mathrm {DC}}+S_{3,\mathrm {GW}}+N_{3,\mathrm {SL}}+N_{3,r}+N_{3,\mathrm {mode}}+N_{3,\mathrm {shot}}
	\nonumber\\
	=&P_{1,\mathrm {DC}}-S_{1,\mathrm {GW}}+N_{1,\mathrm {SL}}+N_{1,r}+N_{1,\mathrm {mode}}+N_{3,\mathrm {shot}}.
\end{align}
In obtaining the second line, we have used the two relations
\begin{equation}
	\alpha_+=-\alpha_-^*,\qquad	n_+	=n_-^*.
\end{equation}
In particular,
\begin{equation}
	S_{3,\mathrm {GW}}=\frac{rP_{\mathrm {BS}}}{8}	\operatorname{Re}(\alpha_-)=-\frac{rP_{\mathrm {BS}}}{8}
	\operatorname{Re}(\alpha_+)=-S_{1,\mathrm {GW}},
\end{equation}
whereas
\begin{equation}
	N_{3,\mathrm {mode}}=\frac{rP_{\mathrm {BS}}}{16}\operatorname{Re}(n_-)=\frac{rP_{\mathrm {BS}}}{16}
	\operatorname{Re}(n_+)=N_{1,\mathrm {mode}}.
\end{equation}
Comparison with $\widetilde M_1$ in Eq.~\eqref{eq.noisyM1} shows that $P_{1,\mathrm {DC}}$, $N_{1,\mathrm {SL}}$, $N_{1,r}$, and $N_{1,\mathrm {mode}}$ enter the two outputs with the same sign. These terms therefore constitute the common-mode contributions of the detector pair \cite{kumar2012versatile}. In contrast, the GW signal enters the two outputs with opposite signs and therefore constitutes a differential-mode contribution. The balanced output is formed by subtracting the two detector outputs \cite{kumar2012versatile,Jin:15},
\begin{equation}
	\widetilde M_{1,3}=\widetilde M_1-\widetilde M_3.
\end{equation}
For ideal balanced detection, the common-mode contributions cancel, whereas the differential GW signal is enhanced by a factor of $2$. The balanced output therefore becomes
\begin{equation}
	\widetilde M_{1,3}=2S_{1,\mathrm {GW}}+N_{1-3,\mathrm {shot}},\label{eq.m13}
\end{equation}
where
\begin{equation}
	N_{1-3,\mathrm {shot}}=N_{1,\mathrm{shot}}-N_{3,\mathrm {shot}}
\end{equation}
denotes the residual shot noise after subtraction. Unlike the common-mode technical noises, the shot-noise fluctuations generated at the two photodetectors are statistically independent and therefore do not cancel. Their PSDs add under subtraction, so the corresponding ASD is
\begin{align}
	A_{{\mathrm {shot}},1-3}=&\left[A_{{\mathrm {shot}},1}^2+A_{{\mathrm {shot}},3}^2\right]^{1/2}\nonumber\\
	=&\sqrt{2} A_{\mathrm {shot}},
\end{align}
where $A_{{\mathrm {shot}},1}=A_{{\mathrm {shot}},3}=A_{\mathrm {shot}}$. Since the GW response function of the balanced output is $2H_{\mathrm {GW}}$, the corresponding shot-noise-limited strain sensitivity is
\begin{equation}
	S^{1/2}_{h,\mathrm {shot},1-3}=\frac{A_{\mathrm {shot},1-3}}{2\left|H_{\mathrm {GW}}\right|}=
	\frac{1}{\sqrt{2}}S^{1/2}_{h,\mathrm {shot}}.
\end{equation}
The same cancellation occurs for the second detector pair,
\begin{equation}
	\widetilde M_{2,4}=\widetilde M_2-\widetilde M_4=2S_{2,\mathrm {GW}}+N_{2-4,\mathrm {shot}}.
\end{equation}
The two balanced outputs measure the real and imaginary quadratures of the effective GW response, respectively. Combining them therefore yields 
\begin{equation}
	S^{1/2}_{h,\epsilon=0}=S^{1/2}_{h,{\mathrm {shot}},1-3}=
	\frac{1}{\sqrt 2}	S^{1/2}_{h,\mathrm {shot}}.
	\label{eq.perfectBalancedSensitivity}
\end{equation}
Here, $\epsilon=0$ denotes ideal balanced detection, as defined below.

For ideal balanced detection, Eq.~\eqref{eq.m13} contains only the GW signal and shot noise, because the common-mode contributions cancel exactly. In practice, however, perfect common-mode rejection cannot be achieved because the BSs cannot realize an exactly $50{:}50$ power splitting ratio. We therefore consider the effect of small BS imbalances. When the light field from the MZI and the reference laser are combined at BS3, we assume that the power transmissivity and reflectivity are $(1/2+\epsilon_1)$ and $(1/2-\epsilon_1)$, respectively, with a small imbalance $|\epsilon_1|\ll1$. The corresponding field transformation can be written as
\begin{equation}
	\begin{pmatrix}
		\sqrt{\frac{1}{2}+\epsilon_1}&\sqrt{\frac{1}{2}-\epsilon_1}\\
		\sqrt{\frac{1}{2}-\epsilon_1}&-\sqrt{\frac{1}{2}+\epsilon_1}
	\end{pmatrix}
	\approx	\frac{1}{\sqrt{2}}
	\begin{pmatrix}
		1+\epsilon_1 & 1-\epsilon_1	\\
		1-\epsilon_1 & -(1+\epsilon_1)
	\end{pmatrix},
\end{equation}
where terms of order $\epsilon_1^2$ and higher have been neglected. The two states after BS3 are therefore
\begin{align}
	&\frac{1}{\sqrt{2}}
	\begin{pmatrix}
		1+\epsilon_1 & 1-\epsilon_1	\\
		1-\epsilon_1 & -(1+\epsilon_1)
	\end{pmatrix}
	\begin{pmatrix}
		\ket{\psi}\\
		r\ket{l'}
	\end{pmatrix}\nonumber\\
	=&
	\frac{1}{\sqrt{2}}
	\begin{pmatrix}
		(1+\epsilon_1)\ket{\psi}+(1-\epsilon_1)r\ket{l'}\\
		(1-\epsilon_1)\ket{\psi}-(1+\epsilon_1)r\ket{l'}
	\end{pmatrix}.
\end{align}
These two states are subsequently divided by BS4 and BS5, respectively. We similarly introduce small imbalance parameters $\epsilon_2$ and $\epsilon_3$ for BS4 and BS5. To first order in the imbalance parameters,
\begin{align}
	&\frac{1}{\sqrt{2}}
	\begin{pmatrix}
		1+\epsilon_2 & 1-\epsilon_2\\
		1-\epsilon_2 & -(1+\epsilon_2)
	\end{pmatrix}
	\frac{1}{\sqrt{2}}
	\begin{pmatrix}
		(1+\epsilon_1)\ket{\psi}+(1-\epsilon_1)r\ket{l'}\\0
	\end{pmatrix}\nonumber\\
	=&\begin{pmatrix}
		\ket{\psi_1}\\\ket{\psi_2}
	\end{pmatrix},
\end{align}
and
\begin{align}
	&\frac{1}{\sqrt{2}}
	\begin{pmatrix}
		1+\epsilon_3 & 1-\epsilon_3\\
		1-\epsilon_3 & -(1+\epsilon_3)
	\end{pmatrix}
	\frac{1}{\sqrt{2}}
	\begin{pmatrix}
		(1-\epsilon_1)\ket{\psi}-(1+\epsilon_1)r\ket{l'}\\0
	\end{pmatrix}\nonumber\\
	=&\begin{pmatrix}
		\ket{\psi_3}\\\ket{\psi_4}
	\end{pmatrix},
\end{align}
where
\begin{align}
	\ket{\psi_1}=&\frac{1}{2}\left[(1+\epsilon_1+\epsilon_2)\ket{\psi}+	(1-\epsilon_1+\epsilon_2)r\ket{l'}\right],\\
	\ket{\psi_2}=&\frac{1}{2}\left[(1+\epsilon_1-\epsilon_2)\ket{\psi}+
	(1-\epsilon_1-\epsilon_2)r\ket{l'}\right],\\
	\ket{\psi_3}=&\frac{1}{2}\left[(1-\epsilon_1+\epsilon_3)\ket{\psi}-
	(1+\epsilon_1+\epsilon_3)r\ket{l'}\right],\\
	\ket{\psi_4}=&\frac{1}{2}\left[(1-\epsilon_1-\epsilon_3)\ket{\psi}-
	(1+\epsilon_1-\epsilon_3)r\ket{l'}\right].
\end{align}
Using these states in the four projection measurements in Eqs.~\eqref{eq.MP1}--\eqref{eq.MP4}, the first output becomes
\begin{align}
	\widetilde M_1=&\frac{P_{\mathrm {BS}}(1+\delta_{\mathrm {SL}})}{2}
	\Bigg|\bra{\Phi_1}\frac{1}{2\sqrt{2}}\Bigg[\sqrt{2}(1-\epsilon_1+\epsilon_2)
	r(1+\delta_r)\nonumber \\ &\times e^{i\delta_{\phi_r}}\ket{l'}+(1+\epsilon_1+\epsilon_2)
	\Big[(\alpha_++n_{l,+})\ket{l+1}\nonumber \\&+(\alpha_-+n_{l,-})\ket{l-1}+(\alpha_++n_{-l,+})\ket{-l+1}\nonumber \\&+
	(\alpha_-+n_{-l,-})\ket{-l-1}\Big]\Bigg]\Bigg|^2+\delta_{1,\mathrm {shot}}\nonumber\\
	\approx&(1-2\epsilon_1+2\epsilon_2)\left(P_{1,\mathrm {DC}}+N_{1,\mathrm {SL}}+N_{1,r}\right)+
	(1+2\epsilon_2)\nonumber \\&\times \left(S_{1,\mathrm {GW}}+N_{1,\mathrm {mode}}\right)+N_{1,\mathrm {shot}}.
\end{align}
Similarly, the other three outputs are
\begin{align}
	\widetilde M_2\approx&(1-2\epsilon_1-2\epsilon_2)
	\left(P_{1,\mathrm {DC}}+N_{1,\mathrm {SL}}+N_{1,r}\right)+(1-2\epsilon_2)\nonumber \\&\times 
	\left(S_{2,\mathrm {GW}}+N_{2,\mathrm {mode}}\right)+N_{2,\mathrm {shot}},\\
	\widetilde M_3\approx&(1+2\epsilon_1+2\epsilon_3)
	\left(P_{1,\mathrm {DC}}+N_{1,\mathrm {SL}}+N_{1,r}\right)+(1+2\epsilon_3)\nonumber \\&\times
	\left(S_{3,\mathrm {GW}}+N_{3,\mathrm {mode}}\right)+N_{3,\mathrm {shot}},\\
	\widetilde M_4\approx&(1+2\epsilon_1-2\epsilon_3)
	\left(P_{1,\mathrm {DC}}+N_{1,\mathrm {SL}}+N_{1,r}\right)+(1-2\epsilon_3)\nonumber \\&\times
	\left(S_{4,\mathrm {GW}}+N_{4,\mathrm {mode}}\right)+N_{4,\mathrm {shot}}.
\end{align}
Using $S_{3,\mathrm {GW}}=-S_{1,\mathrm {GW}}$, $S_{4,\mathrm {GW}}=-S_{2,\mathrm {GW}}$, $N_{3,\mathrm {mode}}=N_{1,\mathrm {mode}}$, and $N_{4,\mathrm {mode}}=N_{2,\mathrm {mode}}$, the two balanced outputs become
\begin{align}
	\widetilde M_{1,3}=&\widetilde M_1-\widetilde M_3\nonumber\\
	=&(-4\epsilon_1+2\epsilon_2-2\epsilon_3)\left(P_{1,\mathrm {DC}}+N_{1,\mathrm {SL}}+N_{1,r}\right)	\nonumber\\
	&+(2\epsilon_2-2\epsilon_3)N_{1,\mathrm {mode}}+2(1+\epsilon_2+\epsilon_3)S_{1,\mathrm {GW}}\nonumber \\
	&+N_{1-3,\mathrm {shot}},\\
	\widetilde M_{2,4}=&\widetilde M_2-\widetilde M_4\nonumber\\
	=&(-4\epsilon_1-2\epsilon_2+2\epsilon_3)\left(P_{1,\mathrm {DC}}+N_{1,\mathrm {SL}}+N_{1,r}\right)
	\nonumber\\
	&+(2\epsilon_3-2\epsilon_2)N_{2,\mathrm {mode}}+2(1-\epsilon_2-\epsilon_3)S_{2,\mathrm {GW}}\nonumber \\
	&+N_{2-4,\mathrm {shot}}.
\end{align}
To characterize the residual common-mode coupling with a single parameter, we define the imbalance parameter as
\begin{align}
	\epsilon=\max\Big(&|-4\epsilon_1+2\epsilon_2-2\epsilon_3|,\nonumber\\
	&|-4\epsilon_1-2\epsilon_2+2\epsilon_3|,|2\epsilon_2-2\epsilon_3|\Big).
\end{align}
Within the small-imbalance approximation, $\epsilon\ll1$. Neglecting the small corrections $\epsilon_2+\epsilon_3$ to the GW response function and using $\epsilon$ as a conservative upper bound on the residual technical-noise coupling, the two balanced outputs can be written schematically as
\begin{align}
	\widetilde M_{1,3}\approx&\epsilon\left(P_{1,\mathrm {DC}}+N_{1,\mathrm {SL}}+N_{1,r}+N_{1,\mathrm {mode}}	\right)+2S_{1,\mathrm {GW}}\nonumber \\
	&+N_{1-3,\mathrm {shot}},\label{eq.balancedM13}\\
	\widetilde M_{2,4}\approx&\epsilon\left(P_{1,\mathrm {DC}}+N_{1,\mathrm {SL}}+N_{1,r}+N_{2,\mathrm {mode}}
	\right)+2S_{2,\mathrm {GW}}\nonumber \\
	&+N_{2-4,\mathrm {shot}}.\label{eq.balancedM24}
\end{align}
Thus, the GW signal is enhanced by approximately a factor of $2$, whereas the common-mode technical-noise amplitudes are suppressed by the imbalance parameter $\epsilon$. The shot-noise contributions remain statistically independent and therefore do not undergo common-mode cancellation. After combining the two balanced outputs, the total equivalent strain sensitivity is
\begin{align}
	S^{1/2}_{h,\epsilon}=&
	\Bigg[\left(\frac{\epsilon}{2}S^{1/2}_{h,\mathrm {SL}}\right)^2
	+\left(\frac{\epsilon}{2}S^{1/2}_{h,r}\right)^2
	+\left(\frac{\epsilon}{2}S^{1/2}_{h,\theta}\right)^2
	+\left(\frac{\epsilon}{2}S^{1/2}_{h,\mathrm {D}}\right)^2 \nonumber \\
	&+\left(\frac{\epsilon}{2}S^{1/2}_{h,\mathrm {P}}\right)^2
	+\frac{1}{2}\left(S^{1/2}_{h,\mathrm {shot}}\right)^2\Bigg]^{1/2}\nonumber\\
	=&
	\frac{1}{2}
	\Bigg \{\epsilon^2\Bigg[
		\left(S^{1/2}_{h,\mathrm {SL}}\right)^2+\left(S^{1/2}_{h,r}\right)^2
		+\left(S^{1/2}_{h,\theta}\right)^2+\left(S^{1/2}_{h,\mathrm {D}}\right)^2\nonumber \\
		&+\left(S^{1/2}_{h,\mathrm {P}}\right)^2\Bigg]+2\left(S^{1/2}_{h,\mathrm{shot}}\right)^2
		\Bigg \}^{1/2}.
	\label{eq.balancedSensitivity}
\end{align}
For ideal balanced detection, $\epsilon=0$, and Eq.~\eqref{eq.balancedSensitivity} reduces to Eq.~\eqref{eq.perfectBalancedSensitivity}. For comparison with the configuration without balanced detection, we may formally assign $\epsilon=2$, for which Eq.~\eqref{eq.balancedSensitivity} reproduces Eq.~\eqref{eq.noBalancedSensitivity}. The value $\epsilon=2$ should be understood as a formal value corresponding to zero common-mode rejection, rather than as a physical BS imbalance within the perturbative regime $|\epsilon_i|\ll1$.

For a balanced-detection output $P_{\mathrm {bal}}$, the response can be decomposed into common-mode and differential-mode contributions as
\begin{equation}
	P_{\mathrm {bal}}=G_{\mathrm {cm}}P_{\mathrm {cm}}+G_{\mathrm {diff}}P_{\mathrm {diff}},
\end{equation}
where $G_{\mathrm {cm}}$ and $G_{\mathrm {diff}}$ denote the response coefficients to the common-mode and differential-mode components, respectively. The ability of the balanced detector to suppress the common-mode contribution relative to the differential signal can be quantified by the common-mode rejection ratio (CMRR), defined in decibels as
\begin{equation}
	{\mathrm {CMRR}}=20\log_{10}\left|\frac{G_{\mathrm {diff}}}{G_{\mathrm {cm}}}\right|.
\end{equation}
From Eqs.~\eqref{eq.balancedM13} and \eqref{eq.balancedM24}, we identify
\begin{equation}
	G_{\mathrm {cm}}=\epsilon,\qquad G_{\mathrm {diff}}=2.
\end{equation}
The imbalance parameter $\epsilon$ is therefore related to the CMRR by
\begin{equation}
	\epsilon=2\times10^{-{\mathrm {CMRR}}/20}.
\end{equation}
References~\cite{8444971,Jin:15,Bruynsteen:21,Jia_2023} report CMRRs of $57.9~{\mathrm {dB}}$, $75.2~{\mathrm {dB}}$, $80~{\mathrm {dB}}$, and $86.9~{\mathrm {dB}}$, respectively. These values correspond to imbalance parameters
\begin{equation}
	\epsilon=2.5\times10^{-3},\quad	3.5\times10^{-4},\quad 2.0\times10^{-4},\quad	9.0\times10^{-5},
\end{equation}
respectively.

\begin{figure*}[!tbhp]
	\centering
	\subfloat[][$\epsilon=2$]{%
		\includegraphics[width=0.48\linewidth]{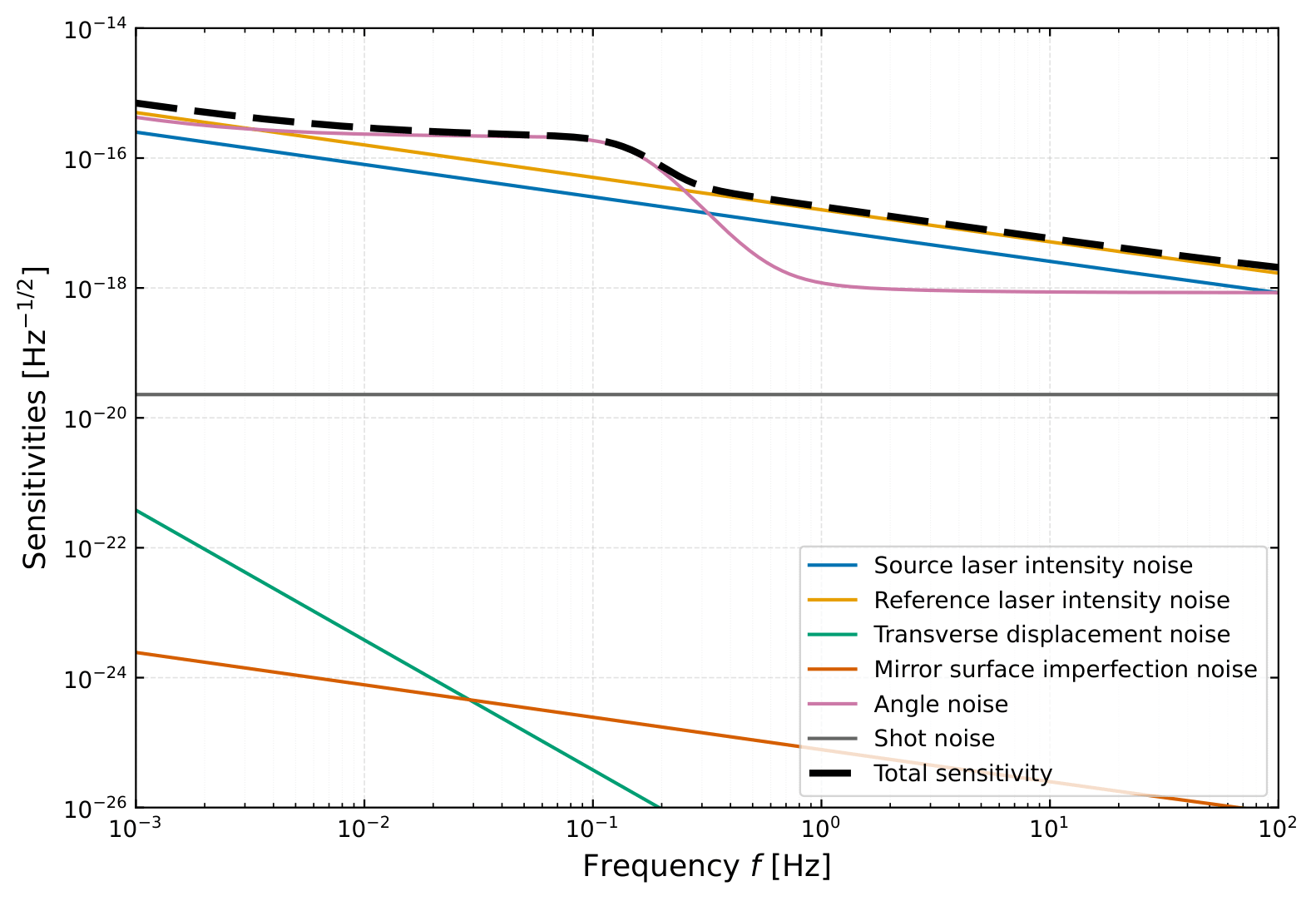}%
		\label{fig.Sensitivity-SA-non}
	}\hfill
	\subfloat[][$\epsilon=10^{-3}$]{%
		\includegraphics[width=0.48\linewidth]{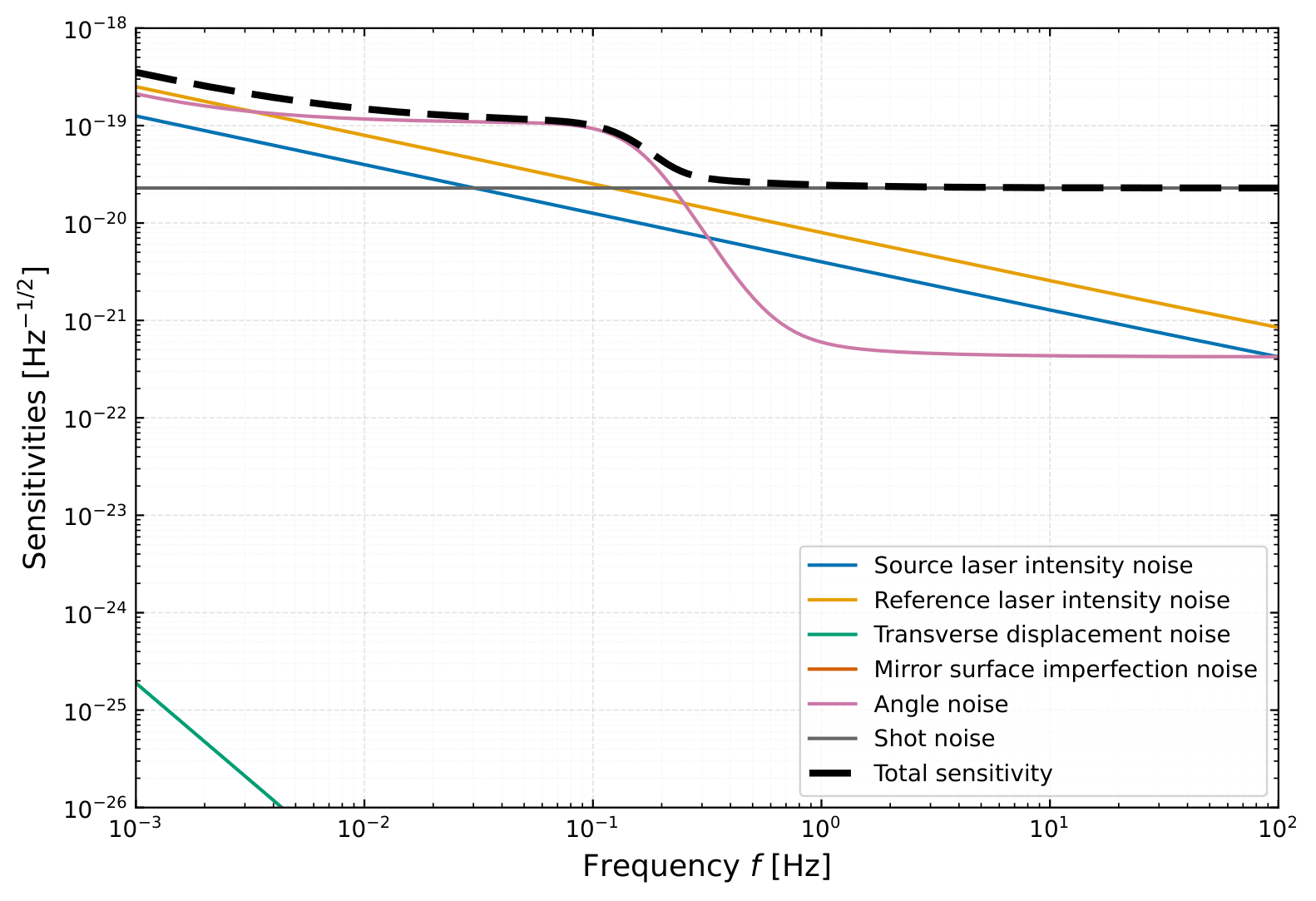}%
		\label{fig.Sensitivity-SA-3}
	}\hfill
	\subfloat[][$\epsilon=10^{-5}$]{%
		\includegraphics[width=0.48\linewidth]{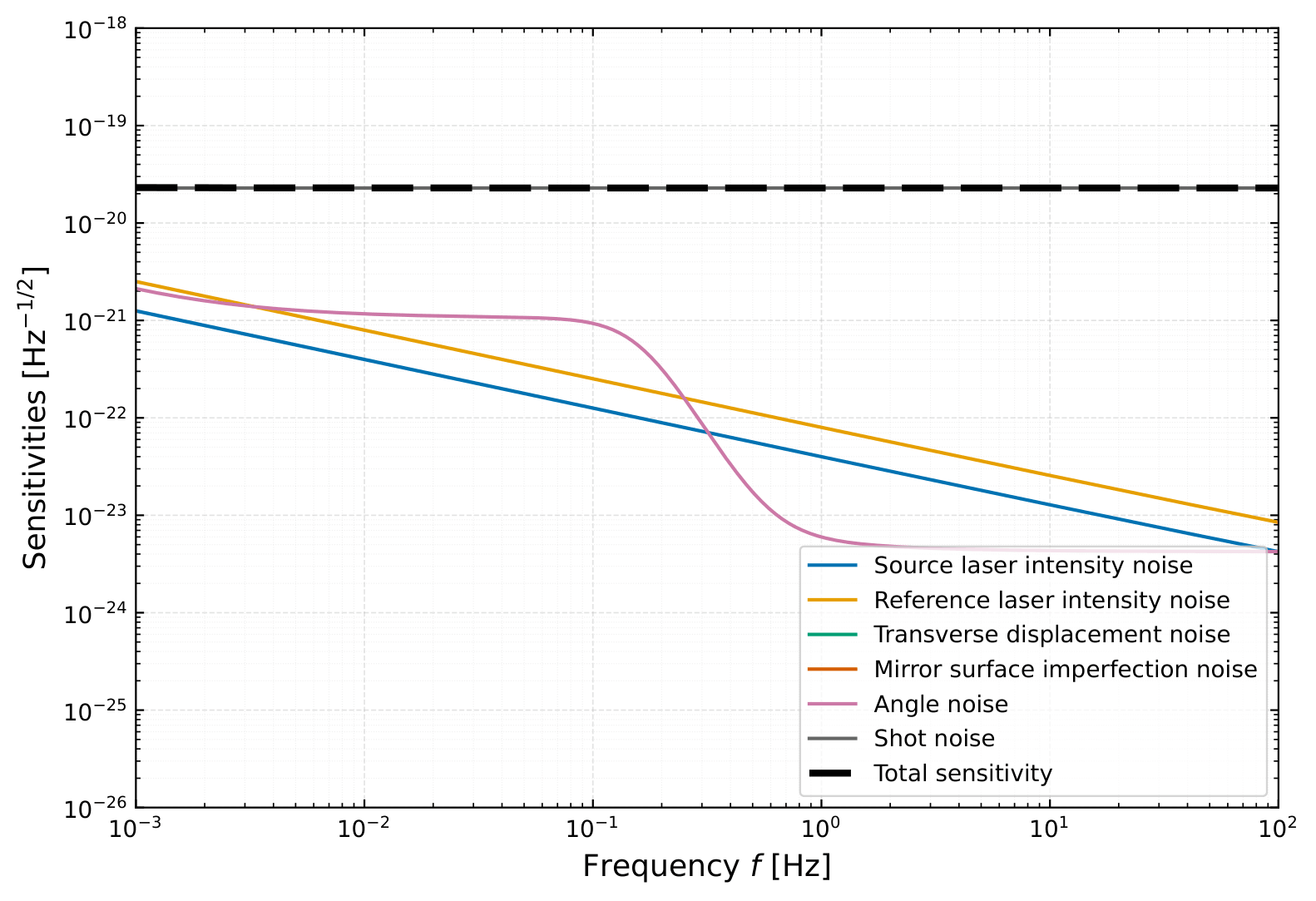}%
		\label{fig.Sensitivity-SA-5}
	}\hfill
	\subfloat[][$\epsilon=0$]{%
		\includegraphics[width=0.48\linewidth]{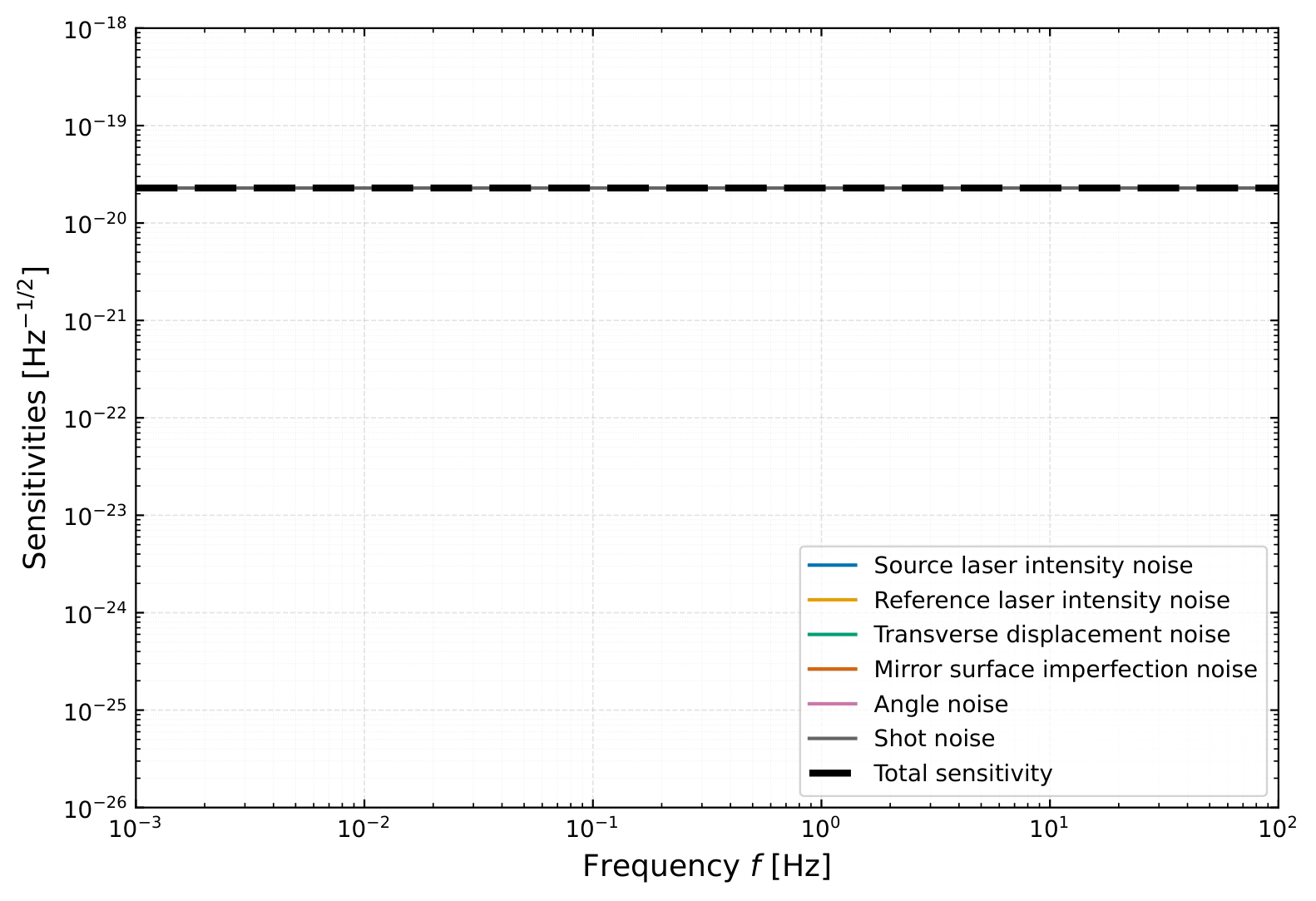}%
		\label{fig.Sensitivity-SA-perfect}
	}\hfill
	\caption{Strain-sensitivity curves for the adopted configuration with $L=4~\mathrm{m}$ and $\gamma=0.1$ for (a) $\epsilon=2$, corresponding to no balanced detection; (b) $\epsilon=10^{-3}$; (c) $\epsilon=10^{-5}$; and (d) $\epsilon=0$, corresponding to ideal balanced detection.}
	\label{fig.SensitivitySA}
\end{figure*}

Figure~\ref{fig.SensitivitySA} shows the sensitivity curves for the adopted configuration with
$L=4~\mathrm{m}$ and $\gamma=0.1$. We first consider the case without balanced detection, corresponding to $\epsilon=2$ in Fig.~\ref{fig.Sensitivity-SA-non}. As discussed above, the compact geometry suppresses the equivalent angular-noise strain sensitivity under the fixed beam waist constraint, while preserving the low-frequency GW response and shot-noise floor. As shown in Fig.~\ref{fig.Sensitivity-SA-non}, angular noise dominates the total sensitivity at
low frequencies. Its contribution decreases rapidly above approximately $0.1~\mathrm{Hz}$ and becomes comparable to the laser-intensity-noise contributions at frequencies of a few tenths of a hertz. At higher frequencies, the reference-laser intensity noise becomes the dominant technical-noise contribution. The total sensitivity decreases from approximately $10^{-16}/\sqrt{\mathrm{Hz}}$ near $1~\mathrm{mHz}$ to the order of $10^{-18}/\sqrt{\mathrm{Hz}}$ near $100~\mathrm{Hz}$. The mirror-surface-imperfection and transverse-displacement contributions remain well below the dominant noise sources and therefore have little influence on the total sensitivity. For balanced detection with $\epsilon=10^{-3}$, shown in Fig.~\ref{fig.Sensitivity-SA-3}, the technical-noise contributions are suppressed by the factor $\epsilon/2=5\times10^{-4}$. Angular noise remains dominant at the lowest frequencies, but decreases rapidly with increasing frequency. The total sensitivity reaches the shot-noise floor at a frequency of order $1~\mathrm{Hz}$ and remains shot-noise limited over most of the higher-frequency range. The sensitivity therefore decreases from approximately $10^{-19}/\sqrt{\mathrm{Hz}}$ near $1~\mathrm{mHz}$ to $2.3\times10^{-20}/\sqrt{\mathrm{Hz}}$ in the shot-noise-limited regime. For $\epsilon=10^{-5}$, shown in Fig.~\ref{fig.Sensitivity-SA-5}, the technical noises are further suppressed by the factor $\epsilon/2=5\times10^{-6}$. All of the technical-noise contributions lie below the shot-noise floor throughout the frequency range considered, so the total sensitivity is essentially shot-noise limited from $1~\mathrm{mHz}$ to $100~\mathrm{Hz}$. Further improvement in the common-mode rejection therefore produces little additional gain in sensitivity. For ideal balanced detection with $\epsilon=0$, shown in Fig.~\ref{fig.Sensitivity-SA-perfect}, all common-mode technical-noise contributions vanish within the present model, leaving the total sensitivity entirely determined by shot noise. The resulting sensitivity is $2.3\times10^{-20}/\sqrt{\mathrm{Hz}}$ throughout the frequency range. The pole frequency of this configuration is $f_p=7.46\times10^4~\mathrm{Hz}$, which is far above the upper frequency of $100~\mathrm{Hz}$ considered here. Therefore, no appreciable attenuation of the GW-induced response occurs within the plotted band.

Finally, we emphasize that the effectiveness of the balanced-detection scheme relies heavily on the different symmetry relations satisfied by $\alpha_\pm$ and $n_\pm$. From the GW-induced transition amplitudes in Eq.~\eqref{eq.alpha}, one finds
\begin{equation}
	\alpha_+=-\alpha_-^*.
\end{equation}
Meanwhile, for the mode-crosstalk contributions induced by angular noise, transverse displacement noise, and mirror-surface-imperfection noise, we obtain the symmetry relation
\begin{equation}
	n_+=n_-^*,
\end{equation}
as given in Eq.~\eqref{eq.noiseCoefficientRelation}. These two relations ensure that the mode-impurity noise, together with the source-laser intensity noise, reference-laser intensity noise, and DC contribution, enters the two detector outputs as common-mode contributions, whereas the GW-induced signal enters as a differential-mode contribution. Balanced detection can therefore suppress these common-mode technical noises while retaining and enhancing the GW signal. The symmetry relation $n_+=n_-^*$ is not generally satisfied by the individual coefficients $n_{l,\pm}$. In particular, $n_{l,\pm}$ contain terms proportional to the OAM index $l$. One way to eliminate these terms is to use an input Bessel mode with $l=0$, as considered in our previous work. However, the $l=0$ Bessel mode has a strong on-axis intensity maximum, which can lead to enhanced localized optical absorption and thermal loading near the center of the mirrors. This behavior follows from the radial dependence of the Bessel modes. For example, $|J_0(k_\perp\rho)|^2$ reaches its maximum value of $1$ at $\rho=0$, whereas $|J_1(k_\perp\rho)|^2$ reaches a maximum of approximately $0.34$ at $\rho\simeq1.84/k_\perp$, and $|J_2(k_\perp\rho)|^2$ reaches a maximum of approximately $0.24$ at $\rho\simeq3.05/k_\perp$. Higher-order modes therefore redistribute the optical power away from the beam center and can reduce localized thermal loading. Related advantages of higher-order Laguerre--Gaussian modes for reducing thermal-noise effects have been proposed and demonstrated in Refs.~\cite{PhysRevD.79.122002,PhysRevLett.105.231102}. In the present work, we instead prepare the input field in the symmetric OAM superposition
\begin{equation}
	\frac{\ket{l}+\ket{-l}}{\sqrt{2}}.
\end{equation}
For this state, the mode-crosstalk coefficients associated with $\ket{l}$ and $\ket{-l}$ contain contributions proportional to $l$ and $-l$, respectively. These terms cancel in the combinations $n_\pm=n_{l,\pm}+n_{-l,\pm}$, yielding the required symmetry relation $n_+=n_-^*$ while allowing the use of nonzero OAM. We further require $l\geq2$. If $l=1$, both components of the input state can couple to the same OAM state $\ket{0}$: the transition $\ket{1}\rightarrow\ket{0}$ belongs to the ``$-$" channel, whereas $\ket{-1}\rightarrow\ket{0}$ belongs to the ``$+$" channel. The corresponding GW-induced amplitudes $\alpha_+$ and $\alpha_-$, as well as the noise amplitudes $n_+$ and $n_-$, therefore contribute to the same output state and can no longer be separated by the projection measurements assumed above. This degeneracy prevents the balanced-detection scheme from exploiting the two symmetry relations in the manner derived here. We therefore restrict the input OAM index to $l\geq2$.

\section{Estimation of detectability}

To estimate the detectability of the proposed detector, we generate frequency-domain GW waveforms using Bilby \cite{bilby_paper,bilby_doi,lalsuite,2022ApJ...935..167A,2018AJ....156..123A,2013A&A...558A..33A}. For each compact binary system, the waveform is characterized by the two polarization components $\tilde h_+(f_g)$ and $\tilde h_\times(f_g)$, which are subsequently combined with the response of the OAM-based detector introduced above. We consider binaries with source-frame mass
\begin{equation}
	1\,M_\odot\leq M_{\mathrm {src}}\leq 10^6\,M_\odot,
\end{equation}
and redshift
\begin{equation}
	10^{-2}\leq	z \leq	10.
\end{equation}
For the cosmology adopted in our calculation, this redshift interval corresponds to luminosity distances ranging approximately from $44.6~{\mathrm {Mpc}}$ to $1.06\times10^5~{\mathrm {Mpc}}$. The component masses are assumed to be equal, so that $m_{1,\mathrm {src}}=m_{2,\mathrm {src}}=M_{\mathrm {src}}$. Cosmological redshift increases the masses measured in the detector frame according to
\begin{equation}
	m_{i,\mathrm {det}}	=(1+z)m_{i,\mathrm {src}},
\end{equation}
and these detector-frame masses are used to generate the observed waveform. To isolate the dependence on the binary mass and distance, we adopt a simplified source configuration. The component spins are set to zero, together with the spin-tilt and azimuthal angles. The binary inclination is also set to zero, corresponding to a face-on system. This choice avoids introducing additional suppression associated with source orientation and provides a representative estimate of the achievable signal strength for the detector configuration considered here. The propagation direction of the GW relative to the detector is chosen as
\begin{equation}
	\theta=\frac{\pi}{4},\qquad	\varphi	=	0.
\end{equation}
This geometry gives nonvanishing responses to the GW polarizations while avoiding special propagation directions for which the OAM-transition amplitudes may be strongly suppressed. For each pair $(M_{\mathrm {src}},z)$, the generated waveforms $\tilde h_+(f_g)$ and $\tilde h_\times(f_g)$ are used together with the detector sensitivity derived in the preceding sections to evaluate the SNR.

We evaluate the SNR separately for the two balanced-detection channels $\widetilde M_{1,3}$ and $\widetilde M_{2,4}$. These two channels measure different quadratures of the GW-induced OAM transition. By taking the Fourier transformation of Eqs. (\ref{eq.hR}--\ref{eq.hI}), the effective GW strains measured by the two balanced outputs are
\begin{align}
	h_{1,3}(f_g)=& \sin\theta  \left [\sin\varphi \tilde h_\times(f_g) -\cos\varphi  \cos\theta\,\tilde h_+(f_g)\right ]	,\\
	h_{2,4}(f_g)=& -\sin\theta  \left [\cos\varphi \tilde h_\times(f_g) +\sin\varphi  \cos\theta\,\tilde h_+(f_g)\right ].
\end{align}
Thus, $\widetilde M_{1,3}$ and $\widetilde M_{2,4}$ provide complementary measurements of the real and imaginary quadratures of the GW-induced transition amplitude. Using the one-sided noise PSD convention, the SNR of each balanced channel is calculated as
\begin{equation}
	\rho_I	=\left[	4	\int_{f_{\mathrm{min}}}^{10^2\,\mathrm{Hz}}	df_g\,
	\frac{|h_I(f_g)|^2}{\left[S^{1/2}_{h,I,\epsilon}(f_g)\right]^2}\right]^{1/2},
\end{equation}
where $I\in\{(1,3),(2,4)\}$. We restrict the integration to the frequency band over which the detector sensitivity is evaluated and include only the portion of the inspiral occurring during the final observer-frame interval $\tau$ before coalescence. Accordingly, the lower integration limit is
\begin{equation}
	f_{\mathrm{min}}= \max\left[10^{-3}\,\mathrm{Hz},f_g(\tau)\right],
\end{equation}
where $f_g(\tau)$ is the observed GW frequency at an observer-frame time $\tau$ before coalescence. At leading order in the inspiral approximation, this frequency is given by \cite{9780198570745.001.0001}
\begin{equation}
	f_g(\tau)=\frac{1}{\pi} \left(\frac{5}{256\tau}\right)^{3/8}M_c^{-5/8}.\label{eq.timeToCoalescene}
\end{equation}
For the equal-mass compact binaries considered here, $M_c=(1+z)2^{-1/5}M_{\mathrm{src}}$ is the detector-frame chirp mass. We adopt $\tau=1\,\mathrm{yr}$ in this paper. Thus, if $f_g(1\,\mathrm{yr})>10^{-3}\,\mathrm{Hz}$, frequencies below $f_g(1\,\mathrm{yr})$ are excluded because they correspond to an earlier portion of the inspiral outside the observing window. If $f_g(1\,\mathrm{yr})\geq10^2\,\mathrm{Hz}$, the integration interval is empty and the corresponding SNR is set to zero. The strain sensitivity of the corresponding balanced-detection channel $S^{1/2}_{h,I,\epsilon}(f)$ is given by Eq.~\eqref{eq.balancedSensitivity}. Finally, the two balanced outputs provide independent signal channels with complementary GW quadratures. Their individual SNRs are therefore combined in quadrature to obtain the total SNR,
\begin{equation}
	\rho=\left[	\rho_{(1,3)}^2	+	\rho_{(2,4)}^2	\right]^{1/2}.
\end{equation}
We adopt $\rho=8$ as the detection threshold and consider signals with $\rho\geq8$ to be detectable. This procedure accounts for the information carried by both quadratures of the GW-induced OAM response while retaining the noise level appropriate to each balanced-detection channel.

\begin{figure}[tbhp]
	\centering
	\includegraphics[width=\linewidth]{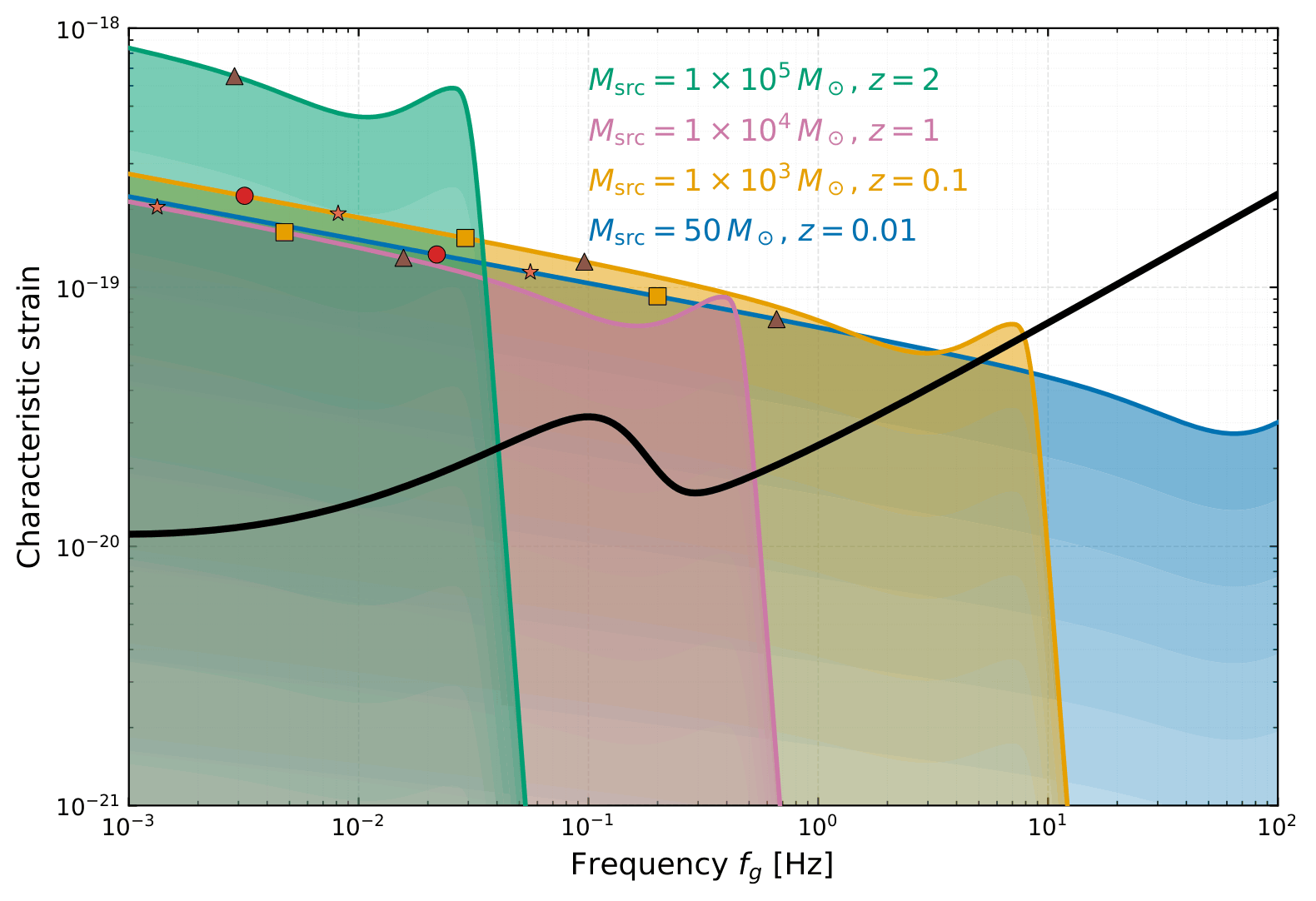}
	\caption{Characteristic strains of representative compact-binary signals and detector noise. The OAM-based configuration (solid black curve) has $L=4~\mathrm{m}$, $\gamma=0.1$, $N=160$, $P_{\mathrm{BS}}=10~\mathrm{kW}$, and $\epsilon=10^{-3}$. The blue, yellow, purple, and green signal curves represent equal-mass BBHs with source-frame component masses and redshifts indicated in the figure. Circles, stars, squares, and triangles on the signal curves mark one year, one month, one day, and one hour before coalescence, respectively, where the corresponding frequencies lie within the plotted range.}
	\label{fig.CharacteristicStrain}
\end{figure}

Before presenting the SNR results, we compare the characteristic strains of the detector noise and several representative GW sources in Fig.~\ref{fig.CharacteristicStrain}. For each noise contribution $n$, the characteristic strain is defined as \cite{moore2014gravitational}
\begin{equation}
	h_n(f)=\sqrt{f}\,S^{1/2}_{h,n}(f),
\end{equation}
where $S^{1/2}_{h,n}(f)$ is the corresponding strain ASD. For a GW signal, we define
\begin{align}
	h_c(f_g)=&2f_g\left [\left | h_{1,3}(f_g)\right |^2+ \left | h_{2,4}(f_g)\right |^2\right ]^{1/2} \nonumber \\
	=&2f_g \sin\theta\left [\cos^2\theta  |\tilde h_+(f_g)|^2+|\tilde h_\times(f_g)|^2 \right ]^{1/2}.
\end{align}
With these definitions, the SNR can be written as \cite{moore2014gravitational}
\begin{equation}
	\rho^2=\int d(\log f_g)\left[\frac{h_c(f_g)}{h_n(f_g)}\right]^2.
\end{equation}
The characteristic-strain representation therefore provides an intuitive visualization of detectability, as a source contributes most strongly to the SNR over the frequency intervals in which its characteristic strain lies above, or is comparable to, the detector-noise curve \cite{moore2014gravitational}. 

Figure~\ref{fig.CharacteristicStrain} compares the characteristic strains of representative compact-binary signals with the characteristic noise strain of the OAM-based detector configuration. Circles, stars, squares, and triangles mark the frequencies corresponding to one year, one month, one day, and one hour before coalescence, respectively. The GW frequency is related to the observer-frame time to coalescence $\tau$ by Eq. \eqref{eq.timeToCoalescene}.

The characteristic noise strain of the OAM-based detector configuration is shown by the solid black curve, with $L=4~\mathrm{m}$, $\gamma=0.1$, $N=160$, $P_{\mathrm{BS}}=10~\mathrm{kW}$, and $\epsilon=10^{-3}$. The blue signal curve represents a stellar-mass binary black hole (BBH) with $M_{\mathrm{src}}=50\,M_\odot$ at $z=0.01$, yielding an SNR of approximately $9.21$, with $f_g(1\,\mathrm{yr})=2\times 10^{-2}\,\mathrm{Hz}$. Its ringdown frequency, approximately $311~\mathrm{Hz}$, lies above the upper limit of the plotted frequency range, so the associated high-frequency roll-off is not shown. The yellow signal curve corresponds to an intermediate-mass BBH with $M_{\mathrm{src}}=10^3\,M_\odot$ at $z=0.1$, yielding an SNR of approximately $22.1$, with $f_g(1\,\mathrm{yr})=3\times 10^{-3}\,\mathrm{Hz}$. Its ringdown frequency is approximately $7.25~\mathrm{Hz}$, beyond which the characteristic strain decreases rapidly. The purple signal curve represents an intermediate-mass BBH with $M_{\mathrm{src}}=10^4\,M_\odot$ at $z=1$, with ringdown near $0.3~\mathrm{Hz}$ and an SNR of approximately $24.7$. Despite its larger mass, its greater luminosity distance reduces the observed strain, which remains below the yellow signal curve over much of their shared frequency range. Finally, the green signal curve corresponds to a massive BBH with $M_{\mathrm{src}}=10^5\,M_\odot$ at $z=2$. Its signal is concentrated at lower frequencies, with ringdown near $3\times10^{-2}~\mathrm{Hz}$. This system produces the largest low-frequency characteristic strain among the four examples and yields the highest SNR, approximately $88.5$, despite its greater distance. These examples illustrate how the signal shifts toward lower frequencies as the redshifted binary mass increases, while the SNR depends on both the signal amplitude and its frequency overlap with the detector sensitivity.


\begin{figure*}[!tbhp]
	\centering
	\subfloat[][$\epsilon=2$]{%
		\includegraphics[width=0.48\linewidth]{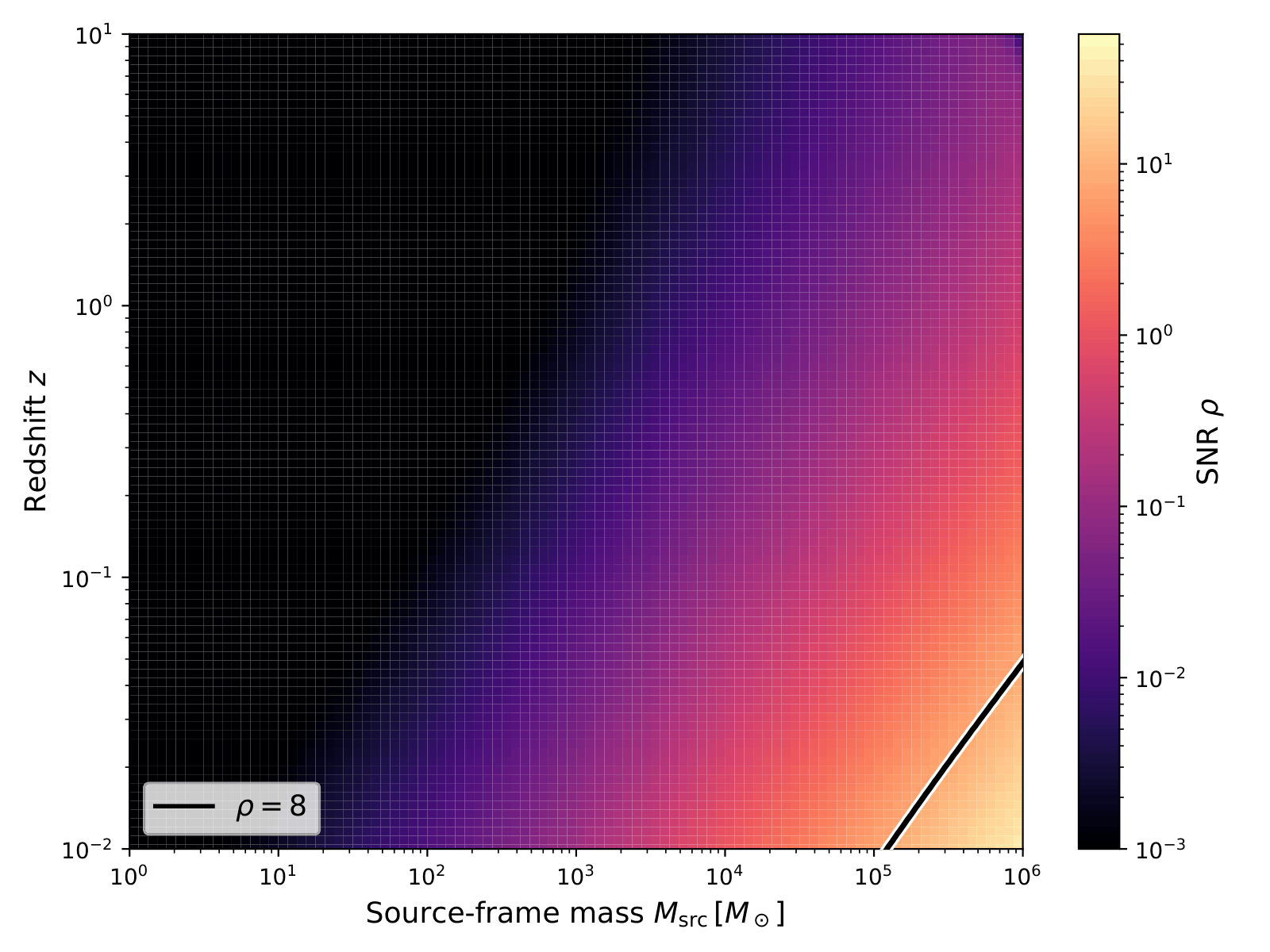}%
		\label{fig.SNR-SA-non}
	}\hfill
	\subfloat[][$\epsilon=10^{-3}$]{%
		\includegraphics[width=0.48\linewidth]{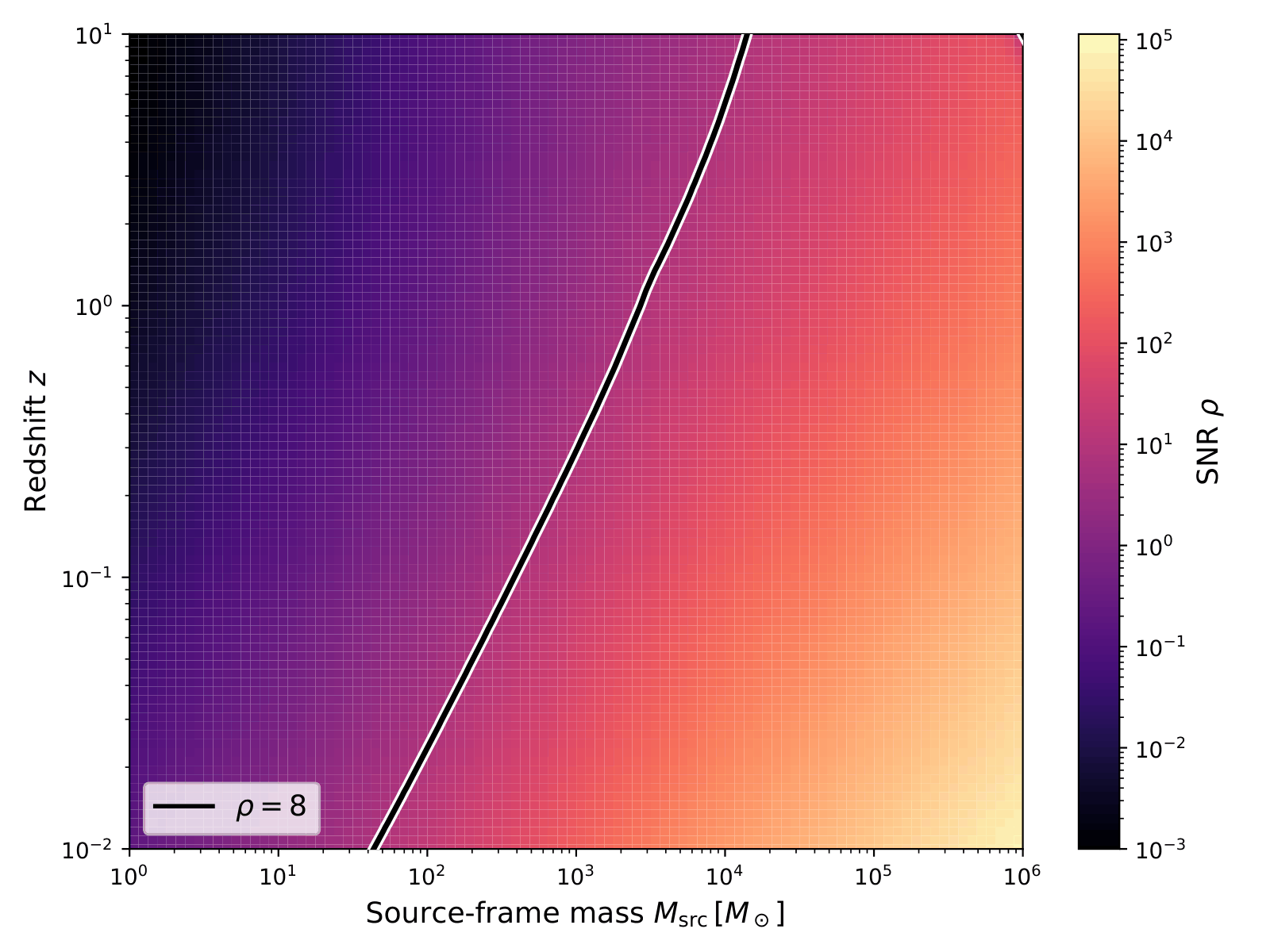}%
		\label{fig.SNR-SA-3}
	}\hfill
	\subfloat[][$\epsilon=10^{-5}$]{%
		\includegraphics[width=0.48\linewidth]{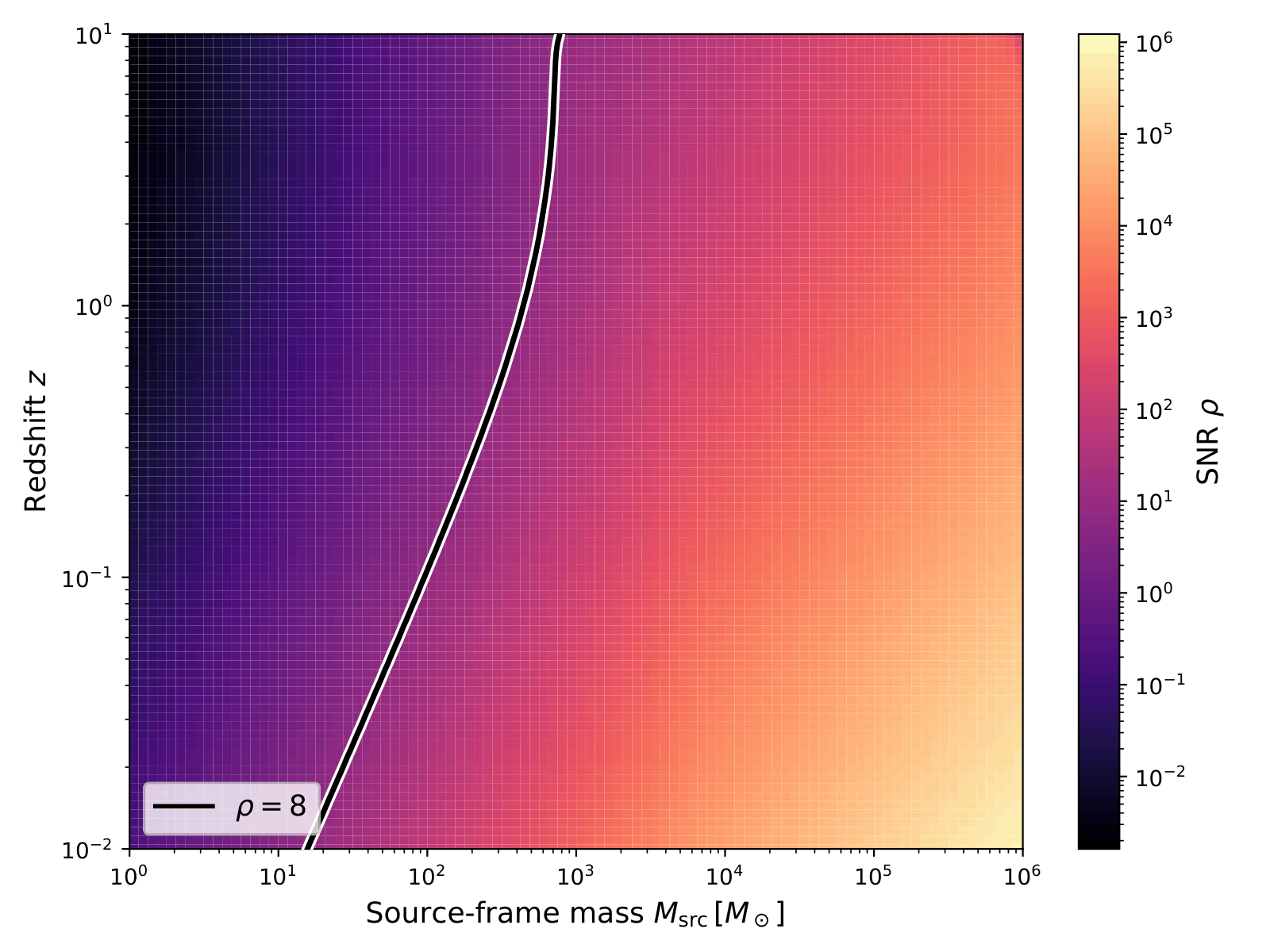}%
		\label{fig.SNR-SA-5}
	}\hfill
	\subfloat[][$\epsilon=0$]{%
		\includegraphics[width=0.48\linewidth]{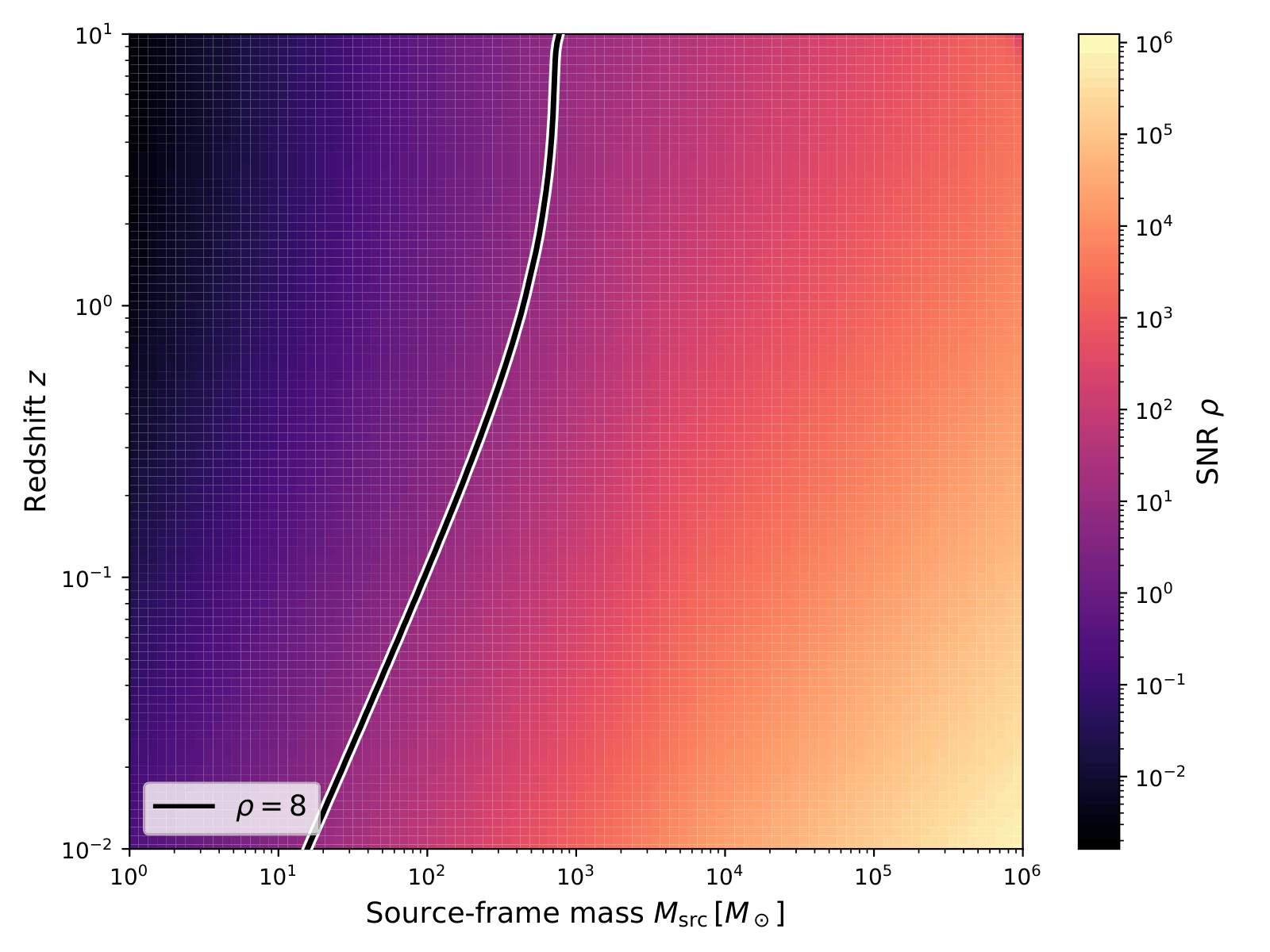}%
		\label{fig.SNR-SA-perfect}
	}\hfill
	\caption{SNR for compact-binary signals in the short-arm configuration with $L=4~\mathrm{m}$ and $\gamma=0.1$ for (a) $\epsilon=2$, corresponding to no balanced detection; (b) $\epsilon=10^{-3}$; (c) $\epsilon=10^{-5}$; and (d) $\epsilon=0$, corresponding to ideal balanced detection.}
	\label{fig.SNRSA}
\end{figure*}

Figure~\ref{fig.SNRSA} shows the SNR for the adopted configuration, with $L=4~\mathrm{m}$ and $\gamma=0.1$. We first consider the case without balanced detection, corresponding to $\epsilon=2$, shown in Fig.~\ref{fig.SNR-SA-non}. In this case, the detector remains strongly limited by technical noise, particularly angular noise at low frequencies. The SNR for a fiducial BBH with $M_{\mathrm{src}}=10^4\,M_\odot$ at $z=1$ is only $0.012$, and the detectable region is confined to very massive and nearby binaries. For example, at $z=0.01$, the detection threshold $\rho=8$ is reached only for source-frame masses of approximately $1.2\times10^5\,M_\odot$ and above. The threshold mass increases to approximately $5.2\times10^5\,M_\odot$ at $z=0.03$ and $1\times10^6\,M_\odot$ at $z=0.05$. At larger redshifts, no source with $M_{\mathrm{src}}\leq10^6\,M_\odot$ reaches the adopted detection threshold. The detectability improves substantially when balanced detection is introduced. For $\epsilon=10^{-3}$, as shown in Fig.~\ref{fig.SNR-SA-3}, the SNR for the fiducial BBH increases to $24.7$, while the maximum SNR reaches the order of $10^5$. The detectable region also expands over several orders of magnitude in both mass and redshift. The threshold source-frame mass is approximately $44\,M_\odot$ at $z=0.01$, $3.8\times10^2\,M_\odot$ at $z=0.1$, $2.7\times10^3\,M_\odot$ at $z=1$, and $1.4\times10^4\,M_\odot$ at $z=10$. Thus, even at the largest redshift considered, binaries with source-frame masses of order $10^4\,M_\odot$ can satisfy $\rho\geq8$. These results demonstrate the importance of suppressing technical noise through balanced detection. Furthermore, as discussed in the angular-noise analysis, the adopted compact configuration also reduces the equivalent strain noise associated with angular noise. For $\epsilon=10^{-5}$, shown in Fig.~\ref{fig.SNR-SA-5}, the technical-noise contributions are further suppressed, substantially expanding the detectable region. The SNR for the fiducial BBH increases to $239$, and the maximum SNR reaches the order of $10^6$. The threshold source-frame mass is approximately $16\,M_\odot$ at $z=0.01$, $95\,M_\odot$ at $z=0.1$, $4.4\times10^2\,M_\odot$ at $z=1$, and $7.8\times10^2\,M_\odot$ at $z=10$. Finally, Fig.~\ref{fig.SNR-SA-perfect} shows the ideal limit $\epsilon=0$, in which common-mode technical noise vanishes within the present model, leaving the detector limited only by shot noise. The resulting $\rho=8$ contour is almost indistinguishable from that for $\epsilon=10^{-5}$, and the SNR for the fiducial BBH increases only slightly, from $239$ to $240$. This close agreement demonstrates that an imbalance parameter of order $10^{-5}$ is sufficient to approach the shot-noise-limited detectability of the detector under the adopted noise model.

\section{Conclusion}

In this work, we investigated the detectability of the previously proposed OAM-based GW detector by incorporating a comprehensive set of relevant noise sources and evaluating the resulting strain sensitivity over the frequency range from $1~\mathrm{mHz}$ to $100~\mathrm{Hz}$. In addition to laser-intensity noise and photodetection shot noise, we considered several mechanisms that induce OAM-state crosstalk, including angular fluctuations of the mirrors, transverse displacement noise, and mirror-surface imperfections. These effects were incorporated into a unified noise model and propagated through the mode-projection and balanced-detection scheme. A central result of our analysis is that the GW-induced transition amplitudes satisfy the symmetry relation
\begin{equation}
	\alpha_+=-\alpha_-^*.
\end{equation}
At the same time, by preparing the incident light field in the symmetric OAM superposition state
\begin{equation}
	\frac{\ket{l}+\ket{-l}}{\sqrt{2}},
\end{equation}
the mode-crosstalk coefficients generated by the technical noises satisfy the complementary relation
\begin{equation}
	n_+	=n_-^*.
\end{equation}
The latter symmetry results from the cancellation of contributions proportional to $l$ and $-l$ between the two components of the input state. The symmetric OAM preparation therefore plays an essential role not only in defining the optical mode structure of the detector, but also in enabling the subsequent rejection of technical noise. These two symmetry relations lead to a clear separation between the GW signal and the dominant technical-noise contributions in the paired detector outputs. The GW-induced terms enter the two outputs with opposite signs and therefore form a differential-mode contribution, whereas the source-laser intensity noise, reference-laser intensity noise, and mode-impurity noise enter with the same sign and behave as common-mode contributions. Subtracting the paired detector outputs consequently enhances the GW signal by approximately a factor of $2$, while suppressing the common-mode technical noises according to the residual imbalance of the balanced-detection system. We characterize this imperfect common-mode rejection through the imbalance parameter $\epsilon$, which is related to the common-mode rejection ratio of the balanced detector. In the ideal limit $\epsilon=0$, the common-mode technical-noise contributions cancel completely within the present model. For finite $\epsilon$, their equivalent strain sensitivities are reduced in proportion to $\epsilon/2$ relative to the corresponding single-output sensitivities. The effectiveness of balanced detection therefore depends critically on both the symmetry of the OAM response and the achievable common-mode rejection. Shot noise behaves differently from the other noise sources considered here. Because the shot-noise fluctuations generated at different photodetectors are statistically independent, they do not cancel when the detector outputs are subtracted. Their PSDs instead add under balanced detection, while the GW signal is doubled. Consequently, shot noise remains after common-mode rejection and ultimately determines the sensitivity floor once the technical noises have been sufficiently suppressed. The transition from a technical-noise-limited regime to a shot-noise-limited regime therefore provides a useful measure of the performance required from the balanced-detection system. 

Another important feature of the proposed detector is the advantage of a compact detector configuration. Under the diffraction-free-range constraint $w_0=\gamma L$, the low-frequency GW-induced transition amplitudes scale as $\alpha_\pm\propto Nw_0k_3$ and are therefore independent of the arm length when $N$ and $w_0$ are fixed. In contrast, the equivalent angular-noise strain sensitivity scales as $S_{h,\theta}^{1/2}\propto(\gamma w_0)^{-1}$. Therefore, a shorter arm allows a larger paraxial parameter and reduces angular-noise coupling. This provides the motivation for adopting $L=4~\mathrm{m}$ and $\gamma=0.1$ in the present analysis.

We further quantify the detectability of the proposed detector by calculating the SNRs of GW signals from compact binary systems over a broad range of source-frame masses and redshifts. In the absence of balanced detection, the detector is strongly limited by technical noise, and only a restricted region of the source parameter space satisfies the detection threshold $\rho\geq8$. Balanced detection substantially improves the detectability. For an imbalance parameter of $\epsilon=10^{-3}$, the maximum SNR reaches the order of $10^5$. Reducing the imbalance parameter to $\epsilon=10^{-5}$ brings the detector close to the shot-noise-limited regime throughout the frequency band considered and increases the maximum SNR to the order of $10^6$. Further improvement in common-mode rejection provides little additional gain once the shot-noise floor dominates.

For the adopted configuration, the cavity pole frequency is $f_p=7.46\times10^4~\mathrm{Hz}$, well above the upper frequency of $100~\mathrm{Hz}$ considered here. The GW-induced response therefore experiences negligible cavity attenuation throughout the frequency band. Together with the reduced angular-noise coupling and common-mode rejection, this makes the compact configuration a promising approach for realizing an OAM-based GW detector.

We emphasize that the noise model adopted in this work is necessarily simplified and should be regarded as a representative estimate rather than a complete noise budget for a specific experimental implementation. Additional noise sources not included in the present analysis may modify the total sensitivity, and correlations among different technical noises may also become relevant in a realistic detector. The results obtained here should therefore be interpreted primarily as an assessment of the potential performance of the OAM-based GW detection scheme under a set of representative noise assumptions. In particular, several of the noise ASDs used in our calculations are strongly dependent on the experimental environment. The angular-noise and transverse-displacement-noise spectra depend on the seismic environment, mechanical isolation, supporting structures, and local disturbances, and can therefore vary significantly between different sites. Likewise, the laser-intensity-noise ASD depends on the specific laser source, stabilization system, and operating power, whereas the mirror-surface-imperfection contribution depends on the substrate quality, coating fabrication, localized absorption, and thermal properties of the optical components. The representative ASDs adopted in this work are based on values reported in the literature and are intended to provide physically motivated estimates over the frequency range of interest rather than universal noise levels. The present analysis also does not include active feedback or feedforward control. In practice, angular motion, beam displacement, laser-intensity fluctuations, and other technical disturbances can be further suppressed using dedicated control systems. The achievable residual noise will depend on the corresponding sensor noise, actuator performance, control bandwidth, and loop design. Therefore, some of the technical-noise levels assumed here may be reduced significantly in an optimized implementation, while other control-related noises may need to be included in a more complete model. The sensitivity and SNR results presented here should therefore be regarded as a first quantitative estimate of the capability of the OAM-based detector, based on representative noise measurements reported in the literature. A more precise assessment of detectability will require a site-specific and component-specific noise budget once the detector location, laser system, optical elements, isolation systems, and control architecture are determined. Direct measurements of the relevant angular, displacement, intensity, and optical-surface noise ASDs under the actual operating conditions will then be necessary to establish a realistic sensitivity curve and to determine the experimentally achievable detection range.

\section*{Acknowledgments}
This work is supported by National Natural Science Foundation of China (12673127 and 125B2103).

\section*{Data availability}
The data that support the findings of this article are not publicly available. The data are available from the authors upon reasonable request.

\end{document}